\documentclass[prx,twocolumn,amsmath,amssymb,floatfix,footinbib,bibnotes]{revtex4-1}
\pdfoutput=1

\usepackage[titletoc]{appendix}
\usepackage[colorlinks=true,citecolor=blue,linkcolor=blue]{hyperref}
\usepackage{amsmath}
\usepackage{graphicx}
\usepackage{dsfont}
\usepackage{changepage}
\usepackage{fancyhdr}
\usepackage{amsthm, amssymb}
\usepackage{floatflt}
\usepackage{float}
\usepackage{array}
\usepackage[usenames,dvipsnames]{color}
\usepackage{slashed}
\usepackage[caption=false]{subfig}

\newcommand{\be}{\begin{align}}
\newcommand{\ee}{\end{align}}

\def \be{\begin{equation}}
\def \ee{\end{equation}}
\def \ba{\begin{array}}
\def \ea{\end{array}}
\def \bea{\begin{eqnarray}}
\def \eea{\end{eqnarray}}
\def \nn{\nonumber}

\def \W{{\Omega}}
\def \e{{\epsilon}}

\def \L{{\Lambda}}
\def \a{{\alpha}}
\def \t{{\theta}}
\def \b{{\beta}}
\def \g{{\gamma}}
\def \D{{\Delta}}
\def \d{{\delta}}
\def \w{{\omega}}
\def \s{{\sigma}}

\def \ve{{\varepsilon}}

\def \G{{\Gamma}}

\def \ba{\begin{align*}}
\def \ea{\end{align*}}

\newcounter{indice}

\def \mrm{\mathrm}

\def \bs{\boldsymbol}
\def \mc{\mathcal}

\def \md{\mathds}

\begin{document}

\title{Pairing from dynamically screened Coulomb repulsion in bismuth   }

\author{Jonathan Ruhman and Patrick A. Lee \\
{  Department of Physics, Massachusetts Institute of Technology, Cambridge, MA 02139 USA  }}
\begin{abstract}
{Recently, Prakash et. al. have discovered bulk superconductivity in single crystals of bismuth, which is a semi metal with extremely low carrier density.
At such low density, we argue that conventional electron-phonon coupling is too weak to be responsible for the binding of electrons into Cooper pairs.
We study a dynamically screened Coulomb interaction with effective attraction generated on the scale of the collective plasma modes.
We model the electronic states in bismuth to include three Dirac pockets with high velocity and one hole pocket with a significantly smaller velocity.
We find a weak coupling instability, which is greatly enhanced by the presence  of the hole pocket.
Therefore, we argue that bismuth is the first material to exhibit superconductivity driven by retardation effects of Coulomb repulsion alone.
By using realistic parameters for bismuth we find that the acoustic plasma mode does not play the central role in pairing. We also discuss a matrix element effect, resulting from the Dirac nature of the conduction band, which may affect $T_c$ in the $s$-wave channel without breaking time-reversal symmetry.    }
\end{abstract}
\maketitle

\section{Introduction}
BCS theory explains how superconductivity emerges from a {\it local} electron-phonon coupling, and leads to the well known expression for the transition temperature
\be
T_c = \Theta_D \, \exp\left[{-{1/ \rho\, V_0}}\right]  \label{BCS}\,.
\ee
Here $\rho$ is the density of states per spin at the Fermi energy, $\Theta_D$ is the Debye temperature and $V_0$ is the phonon mediated interaction.
In metals the local electron-phonon coupling is, indeed, the most important one~\cite{Morel1962}.

However, as pointed out first by Gurevich, Larkin and Firsov (GLF)~\cite{Gurevich1962}, this theory runs into a serious problem when considering superconductivity in dilute metallic systems, such as doped semiconductors and semimetals. The reason is that in three dimensions the density of states, $\rho$, decreases to zero as the density of carriers is decreased. GLF have concluded that in non-ionic crystals the lowest possible density for superconductivity is $n_{GLF} \sim 10^{19} \mrm{cm}^{-3}$~\footnote{This density bound, $n_{GLF}$, is obtained by assuming a given band mass $m$. Then $n_{GLF}$ is proportional to $n_{GLF}\propto m^{-3}$. Thus, systems with small band masses, such as Bi$_2$Se$_3$, may have a significantly higher bound.}.
For ionic crystals, on the other hand, they proposed that coupling to the long range polarization of the longitudinal optical phonon, within the random-phase-approximation (RPA), may lead to superconductivity at densities much lower than $n_{GLF}$. It is, however, important to note that the frequency of the longitudinal mode must still be much smaller than the Fermi energy, $\w_L \ll \e_F$, such that the Coulomb repulsion may be renormalized.

Indeed, most known superconductors with a density below the limit $ n<n_{GLF}$ are ionic compounds~\footnote{Although it should be mentioned that the restriction $\w_L \ll \e_F$ is typically not fulfilled. The extreme example is SrTiO$_3$\cite{Schooley1964,Lin2014,chandra2017prospects}, where the phonon frequency is significantly higher than the Fermi energy. But also in the case where the Fermi energy is greater, such as doped PbTe~\cite{matsushita2006type}, SnTe~\cite{Sasaki2012}, Bi$_2$Se$_3$~\cite{hor2010} and the half-Heuslers, the two energy scales are still of the same order of magnitude. Thus, there is an open question regarding how repulsion gets screened, even in the ionic systems.}.
Examples are the topological half-Heusler semimetals YPtBi, LuPtBi, LaPtBi with densities as low as $n = 2\times 10^{18} \mrm{cm}^{-3}$~\cite{butch2011superconductivity,tafti2013superconductivity,nakajima2015topological,kim2016beyond} and SrTiO$_3$ with a density as low as $n = 5\times 10^{17} \mrm{cm}^{-3}$~\cite{Schooley1964,Koonce1967,Lin20133,Lin2014}.

The recent discovery of superconductivity in bismuth~\cite{Prakash2016}, however, is in direct contradiction to the GLF theory. On the one hand the density of carriers, which is $n = 3\times 10^{17}\,\mrm{cm}^3$, is well below the bound (for a detailed discussion on the irrelevance of phonons and other local interactions to superconductivity see Appendix~\ref{app:phonon}).
On the other hand bismuth is a single element crystal, and therefore has no polar phonon modes. Thus, the puzzle is: What is the source for long range attractive interaction which allows such a low density system to become superconducting?

The discovery of high-$T_c$ superconductivity has raised the possibility of pairing in lightly doped Mott insulators due to strong correlations effect. However, in Bi the bands are predominantly wide $6p$ bands and strong correlation is not expected. (For an alternative viewpoint on pairing from strong correlation effects see Ref.~\cite{baskaran2017theory}). There remain two possible sources for long-ranged attractive interactions: (i) Soft critical fluctuations coming from a nearby critical point (the critical point maybe associated with an instability of the Fermi gas or a structural transition of the ions in the crystal)~\cite{Metlitski2015,Lederer2015,Kozii2015,wang2016Topological}. (ii) The second possibility is collective plasma modes~\cite{Takada1978,Takada1980b,Rietschel1983,Grabowski1984,Takada1992,takada1993s}, which mediate a long ranged attraction. Since there is no experimental evidence of a nearby critical point we focus, in this paper, on the latter. The plasmonic modes are longitudinal collective modes, which are related to longitudinal phonon modes in a polar crystal. Therefore they are, in principle, compatible with the GLF theory. The plasma modes set the energy scale for the retardation (frequency dependent) effect of the screened Coulomb interaction leading to effective attractive pairing channels.

Takada~\cite{Takada1978} was the first to apply the GLF theory to study the possibility of plasmonic superconductivity. He concluded that a superconducting instability can occur if the coupling is strong, i.e. $r_s >1$ (where $r_s$ is proportional to the ratio between Coulomb interaction and kinetic energies). In this limit the plasma frequency is typically higher than the Fermi energy. Thus, both Eliashberg theory and the RPA are out of their limits of validity. For this reason there has been a lot of controversy in the literature whether or not this instability exists even in theory \cite{Rietschel1983,Grabowski1984,takada1993s,Grimaldi1995,Richardson1996}.

In this work we focus on the limit of weak coupling~\cite{Ruhman2016} where RPA is a good approximation. Our goal is to understand the origin of superconductivity in bismuth, where the measured plasma frequency is smaller than the Fermi energy~\cite{Tediosi2007}, indicating that indeed $r_s <1$. We derive the Eliashberg equation which is used to compute $T_c$ numerically in an isotropic Dirac semi-metal band structure, which approximates that of bismuth. It includes three Dirac-electronic pockets and one parabolic-hole pocket~\cite{Wolff1964,Liu1995,Zhu2011}, where the Dirac velocity is significantly larger than the Fermi velocity of the holes.

We find, in contrast to Takada~\cite{Takada1978,Takada1992}, that there exists a weak coupling instability towards superconductivity and discuss the possibility that this instability extends to arbitrarily weak coupling. By calculating $T_c$ with and without the hole pocket we show that the holes drastically enhances $T_c$ in the weak coupling limit.
 Our results indicate that Bi may be the first example of weak coupling superconductivity driven by retardation effects of Coulomb repulsion alone, without the help of phonons. We also note that in this case the isotope effect should be completely absent, as the mass of the bismuth atoms plays no role in the theory.

It is also important to note that in the case of bismuth, where there are multiple Fermi surfaces with different Fermi velocities, there may exist an additional collective mode: the so called {\it acoustic plasmon}~\cite{pines1956electron,ruvalds1981there,bennacer1989acoustic,bennacer1989calculated,Chudzinski2011}.  This mode has been considered as a possible mechanism for attractive interactions and superconductivity in semi-metals already a long time ago~\cite{Garland,RADHAKRISHNAN1965247,pashitskii1968plasmon,frohlich1968superconductivity,ruvalds1981there,entin1984acoustic}. Such a mode can contribute greatly to superconductivity when the velocity ratio is very large (in this limit the slower band becomes equivalent to the "jelly" in the "jellium" model and the acoustic plasmon is just the acoustic phonon). However, we show that for realistic parameters in bismuth a weak coupling instability occurs even in the limit where this mode does not exist, and therefore we argue that the acoustic plasmon is not an essential ingredient for superconductivity in this material. Nevertheless, we find that the hole band tends to enhance the transition temperature.

Finally, in Bi strong spin-orbit coupling leads to nearly massless Dirac conduction bands. Anderson~\cite{anderson1959theory} pointed out that in the presence of spin-orbit-coupling, one pairs time reversed eigenstates  rather than opposite spins, such that the s-wave pairing BCS theory and Eq.~\eqref{BCS} remain valid. We find that in the case where the paired states reside on a Dirac cone, this conclusion is violated due to a matrix element effect, as recently found in quadratic band touching semi-metals~\cite{Savary}. For more details see Appendix \ref{andersons}.

\tableofcontents

\section{Band structure of Bismuth}
We start by quickly reviewing the single particle dispersion of the electron and hole bands.
Bismuth has rhombohedral crystal structure, which belongs to the point group symmetry class $D_{3d}^5$ ($R \bar 3 m$)~\cite{Liu1995}. Strong spin-orbit coupling leads to three electronic pockets centered at the three $L$-points and one hole pocket revolving around the $T$-point.
Each one of the electron pockets is descried by a Dirac Hamiltonian~\cite{Wolff1964}
\be
H_e(\bs k) = \left[v_{z} k_z \s^z + v_{\perp}\left( k_x \s^x + k_y \s^y  \right)\right] s^x + \D_{bg} s^z \,,\label{H_e}
\ee
where $z$ denotes the direction along the $\G-L$ line, and the parameters appear in Table. \ref{tab:parameters}. Here $\s^i$ and $s^i$ are Pauli matrices (we use the same notation as Wolff~\cite{Wolff1964} for the band notation).
Bismuth is time reversal and inversion symmetric. In this basis the symmetries are implemented by $\mc T = i \, \s^y \, s^z$ and $\mc P = s^z$, respectively. The Hamiltonian (\ref{H_e}) is diagonalized by the unitary transformation $\L(k)$, i.e. $\L^\dag (k) H_e(k)\L(k) = \mrm{diag}\left\{ \e_k,\e_k,-\e_k,-\e_k\right\}$. Here $\e_k = \sqrt{\left(v_z k_z \right)^2 + v_\perp^2(k_x^2+k_y^2) + \D_{bg}^2 }$.

The hole band is located around the $T$-point (i.e. the trigonal direction). Here the gap between the conduction and valence bands is much larger than the chemical potential, and therefore the dispersion of these carriers is essentially parabolic
\be
H_h(\bs k) = {k_z ^2 \over 2 M_z} + {k_x ^2 +k_y^2\over 2 M_\perp}\,,
\ee
where the z-direction points along the trigonal direction, $M_z = 0.72 \, m_e$ and $M_\perp = 0.07\, m_e$~\cite{Zhu2011}.

The anisotropy in bismuth is fairly large and may have important implications. However, in this work, we will be interested in addressing the puzzles regarding the appearance of superconductivity in bismuth. Thus, for the sake of simplicity, we will approximate the band structure to be isotropic.
 The isotropic approximation is obtained by considering a mean velocity $v = \left(v_\perp^2 v_z \right)^{1/3} $ for the Dirac electrons and a mean mass term $M = (M_{\perp}M_z )^{1/3}$ for the parabolic holes. This is formally equivalent to redefying the coordinates \cite{Zhu2011} and importantly preserves the density of states at the Fermi level.

The parameters considered in this work for the electron and hole bands appear in Table \ref{tab:parameters}.

\begin {table*}
\caption { List of parameters used in this work. } \label{tab:parameters}
\begin{center}
    \begin{tabular}{| l | l | l | l |}
    \hline
    \textbf{Parameter} & \textbf{Notation}  & \textbf{Value} & \textbf{Reference} \\ \hline
    Density  & $n$ & $3\times 10^{17}\,\mrm{cm}^{-3}$ &\cite{issi1979low,Liu1995}\\ \hline
   Total plasma frequency&$ \w_p$& $ 17\, \mrm{meV}$ & \cite{Tediosi2007}\\\hline
   Total Thomas-Fermi momentum&$ \kappa_{TF}$&  $\sqrt{N q_{TF}^2 + Q_{TF}^2}$ &\\\hline
   Static dielectric constant&$ \ve_\infty $& $ 30 $ (isotropic approx.)& see Eq.~(\ref{alpha}) \\ \hline
   Unit cell volume  & $\mc V_{uc}$ & $64 \, \AA{}^3$ &\cite{Liu1995}\\ \hline
    Debye temperature  & $\Theta_D$ & $118 \, K$ &\cite{desorbo1958low}\\ \hline
    Transition temperature  & $T_c$ & $0.5 \, \mrm{mK}$ & \cite{Prakash2016}\\\hline
     Number of electronic pockets & $N$ & 3&  \cite{issi1979low,Liu1995} \\\hline
    Electron Fermi energy & $\e_F$ & $30\, \mrm{meV}$ &  \cite{issi1979low,Liu1995,Tediosi2007} \\\hline
       Dirac velocities & $v_\perp\,, v_z$ & $ 8.1\times 10^5 \,,6.6\times 10^4  \, \mrm{m/sec}$ &\cite{,issi1979low,Zhu2011}\\ \hline
       Average electronic velocity & $v = \left(v_\perp^2 v_z \right)^{1/3}$ & $3.5\times 10^{5} \, \mrm{m/sec}$ &\\ \hline
       Dirac mass   & $\D_{bg}$ & $7.5\, \mrm{meV}$ &\cite{issi1979low,Zhu2011,Wolff1964}\\ \hline
       Average electron Fermi momentum   & $k_F $ & $1.4\times 10^{8}\, \mrm{m^{-1}}$ &\cite{issi1979low,Zhu2011} \\\hline
       Electron density of states (per pocket per spin)    & $\rho$ & $9.2\times10^{18}\,\mrm{eV^{-1} \,cm^{-3}}$ &\\ \hline
        Electron plasma frequency    & $w_p$ & $\sqrt{4\pi e^2 2N\rho v^2 / 3\ve} $ &\\ \hline
       Electron Thomas-Fermi momentum    & $q_{TF}$ & $\sqrt{4\pi e^2 2N\rho / \ve} $ &\\ \hline
       Hole Fermi energy   & $E_F$ & $8.6\, \mrm{meV}$ &\\ \hline
       Hole masses   & $M_{\perp}\,,M_z$ & $0.07\,, 0.72\, \mrm{m_e}$ &\cite{Liu1995,issi1979low,Zhu2011}\\ \hline
       Average hole mass   & $M = \left( M_\perp^2 M_z\right)^{1/3}$ & $0.15\, \mrm{m_e}$ &\\ \hline
       Average hole Fermi momentum   & $K_F $ & $1.85\times 10^{8}\, \mrm{m^{-1}}$ &\cite{issi1979low,Zhu2011}\\ \hline
       Average hole Fermi velocity   & $V_F$ & $1.4\times 10^{5} \, \mrm{m/sec}$ &\\ \hline
       Hole density of states (per spin)  & $R$ & $3.5\times10^{19}\,\mrm{eV^{-1} \,cm^{-3}}$ &\\ \hline
       Hole plasma frequency    & $W_{p}$ & $\sqrt{4\pi e^2 2R V_F^2/ 3\ve} $ &\\ \hline
       Hole Thomas-Fermi momentum    & $Q_{TF}$ & $\sqrt{4\pi e^2 2R / \ve} $ &\\ \hline
    \end{tabular}
    \end{center}
    \end {table*}

\section{The dynamically screened Coulomb interaction}
We consider the effects of the long-ranged Coulomb interaction
\be
V(i\w,q) = {4 \pi e^2 \over \ve(i\w,q)q^2}\,.\label{Coulomb}
\ee
where $\w$ is a bosonic Matsubara frequency and the dielectric constant is given by
\be
\ve(i\w,q) = \ve_\infty - {4\pi e^2 \over q^2} \left[ \Pi_e(i\w,q) +\Pi_h (i\w,q) \right] \,.\label{epsilon}
\ee
Here $\ve_\infty$ is the {\it static} dielectric constant coming, mainly, from low momentum interband transitions. $\Pi_{e,h}$ are the intraband polarizations of the electrons and holes, respectively, which are calculated within the random-phase-approximation (RPA) [see Eqs.~(\ref{e_polarization},\ref{h_polarization}) in Appendix \ref{app:polarizations}]. Note that here we have neglected the polarization due to interpocket transitions, since it is only important at high momentum.

\subsection{Collective modes}
As explained, in this paper we focus on the possibility that electronic pairing in bismuth comes from the collective electronic modes, which set the scale of retardation effects and open new pairing channels. The collective modes are given by the zeros of the dielectric function (\ref{epsilon}). In bismuth there is a significant difference between the Fermi velocity of the holes and electrons. This leads to the appearance of two longitudinal plasma modes~\cite{pines1956electron,bennacer1989acoustic}:

\begin{figure}
 \begin{center}
    \includegraphics[width=0.9\linewidth]{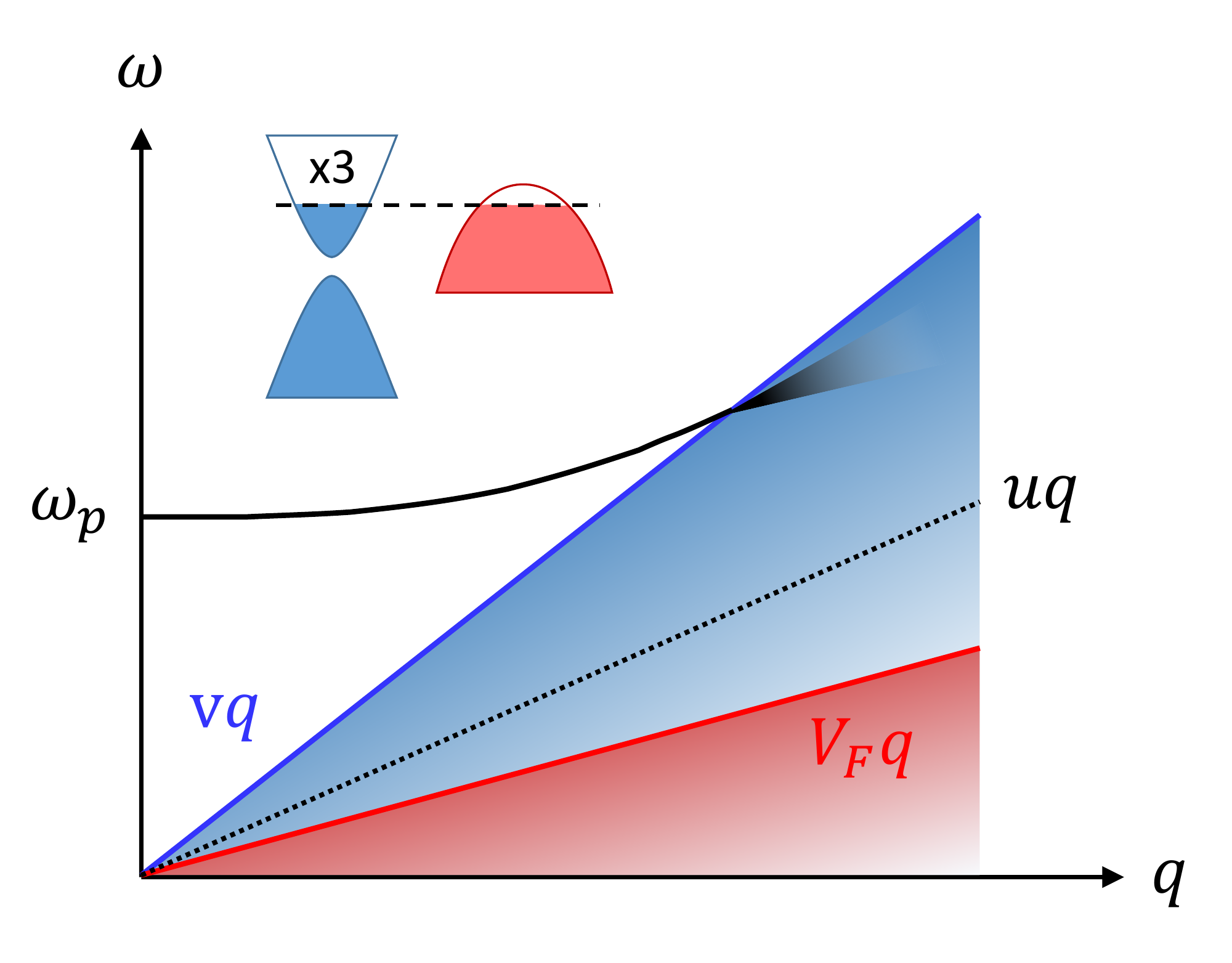}
 \end{center}
\caption{ Schematic representation of the collective modes in bismuth in the limit of small $q$.
The two shaded regions (blue and red) mark the particle continuum of the electrons and holes, respectively. The onset of these regions occurs at the lines defined by $\w = v q$ and $\w = V_F q$, respectively.
The solid black line is the gapped plasma mode, which at $q =q_c\gtrsim \w_p / v$ enters the particle hole continuum of the electron pockets and becomes over damped. The acoustic plasmon (black dotted line) disperses linearly, $\w = u q$, and lies in the particle-hole continuum of the electrons. When the ratio between the velocities is small $\d_v = V_F / v \ll 1$ this mode is weakly damped and can mediate superconductivity. On the other hand as the ratio $\d_ v$ increases, at a certain value of $\d_v < 1$, the acoustic plasmon pole disappears (see Fig.~\ref{fig:ap}).  The inset shows the schematic band structure in bismuth with three slightly doped Dirac pockets and a single hole pocket with larger mass and smaller velocity.   }
 \label{fig:modes}
\end{figure}

\subsubsection{ The gapped plasmon}\label{sec:plasmon}
The first pole is the standard plasmon, which describes a collective compression mode of {\it total} charge. It is obtained in the limit $\w \gg vq,\, V_F q$. In this case both the polarizations of the electrons and the holes in Eq.~(\ref{epsilon}) are in the dynamic regime $\ve(i\w,q) \approx \ve_\infty \left( 1+ \w_p^2 / \w^2 \right)$, where $\w_p = \sqrt{N \,w_p ^2 + W_{p}^2}$ is the total plasma frequency,
\begin{align}
w_p  \equiv {v q_{TF} \over \sqrt{3}}   \;\;\;;&\;\;q_{TF} \equiv \sqrt{8\pi  e^2 \rho\over \ve_\infty}  = \sqrt{ 4 \a \over \pi}k_F\,,\label{wp}\\
W_p  \equiv  {V_F Q_{TF} \over \sqrt{3}} \;\;\;;&\;\;Q_{TF} \equiv \sqrt{8\pi e^2 R\over \ve_\infty } = \sqrt{{4 \a \d_k^2 \over \pi \d_v}}k_F\nn
\end{align}
are the plasma frequencies and Thomas-Fermi momenta of the electrons and holes, respectively, $N$ is the number of electron pockets, $\rho = k_F^2 / 2 \pi^2 v$ and $R = K_F^2 / 2 \pi^2 V_F$ are the electron and hole density of states per spin and pocket, $\d_k = K_F / k_F$ and $\d_v = V_F / v$. Thus the total plasma frequency, $\w_p$, and the total Thomas-Fermi momentum, $\kappa_{TF} = \sqrt{N q_{TF}^2 + Q_{TF}^2}$ can be written as
\be
\w_p = w_p \sqrt{N + \d_k^2 \d_v}\;;\; \kappa_{TF} = q_{TF}\sqrt{N + \d_k^2 / \d_v}\,. \label{kappa}
\ee

The parameter
\be
\a = {e^2 \over \ve_\infty v}\label{alpha}
\ee
 is the effective fine structure constant, which appropriately quantifies the coupling strength in a Dirac dispersion.
Plugging in the average velocity in Table~\ref{tab:parameters} we find that $\a = 6.2/\ve_\infty$. The corresponding plasma frequency taken from Eqs. (\ref{wp} and \ref{kappa}), is given by $\w_p = 2.9 \,v k_F\, / \,\sqrt{\ve_\infty}$. To fix the plasma frequency to be equal to the experimental measurement ($\w_p = 18\,\mrm{meV}$) we set $\ve_\infty  = 30$~\cite{Tediosi2007}, which corresponds to $\a = 0.2$. We note that there is uncertainty in this value because it depends strongly on the assumption of isotropic bands. In reality $\ve_{\infty}$ is estiamted to be higher~\cite{Boyle1960} and it is not clear what is the effective coupling strength in the highly anisotropic bands.

The dispersion of the plasmon mode is schematically presented in fig.~\ref{fig:modes} (solid black line). As can be seen at $q>0$ the mode weakly disperses, until at $q_c \sim  \w_p / v$ it crosses into the p-h continuum of the electron pockets (marked by the shaded blue region in the figure) where it becomes over damped. Therefore, there is a limitation on the phase space for scattering by plasmons, which becomes more important at weak coupling, where $q_c < 2k_F$. The phase space constraint has important implications on superconductivity since it reduces the strength of the attractive interaction coming from the plasmon mode (For more details on the limited phase space of the plasmon see Appendix \ref{app:collective modes}).

\subsubsection{The acoustic plasmon}\label{sec:acoustic_plasmon}
The second pole is the {\it acoustic plasmon}~\cite{pines1956electron,bennacer1989calculated,Chudzinski2011} observed at lower frequency $vq> \w \gtrsim V_F q$.
This mode describes an {\it out-of-phase} compression of the two charged fluids, which does not affect any modulation in the total charge. As a result it is neutral and is thus acoustic, similar to the zero sound mode of a neutral Fermi liquid. The main difference compared to a neutral Fermi liquid is, however, that the mode is damped because it lies in the p-h continuum of the electrons.

Here we also point out that in the limit of $V_F\ll v$, the hole serves as positive charge background and the model becomes the "jellium" model. In this case the acoustic plasmon becomes the sound wave of the jellium liquid~\cite{Ashcroft}. The usual BCS theory of the exchange of the acoustic phonon then applies, where the acoustic plasmon takes the place of the phonon. We will see later that Bi is far from this limit.

To obtain the linear dispersion of the acoustic plasmon we seek the zeroes of (\ref{epsilon}) near $\w,\,q \rightarrow 0$ which are given by the equation
\be
f(z) = - {\nu \over \d_v} f( z / \d_v) \label{acoustic plasmon}
\ee
where $z = \w/vq$ and $f(z) = {z\over 2} \log\left( {z+1 \over z-1} \right)-1$. The solution of equation (\ref{acoustic plasmon}), denoted by $z = z_0$, depends on two parameters: $\nu \equiv \d_k^2 / N $ and $\d_v$ (see discussion below Eq.~(\ref{wp}) for definitions). If there exists a solution it is always in the regime $z<1$ and thus the l.h.s of Eq. (\ref{acoustic plasmon}) contributes a non-trivial imaginary part, which leads to damping of the mode reflected by $\mrm {Im} z_0 < 0$. The general solution can thus be written as
\be
\w = u q = (u_1-i u_2)q \label{solution}
\ee
where $u = v z_0$ and $u_{1,2}$ are real positive numbers.

\begin{figure}[t]%
    \centering
    \subfloat{{\includegraphics[width=0.8\linewidth]{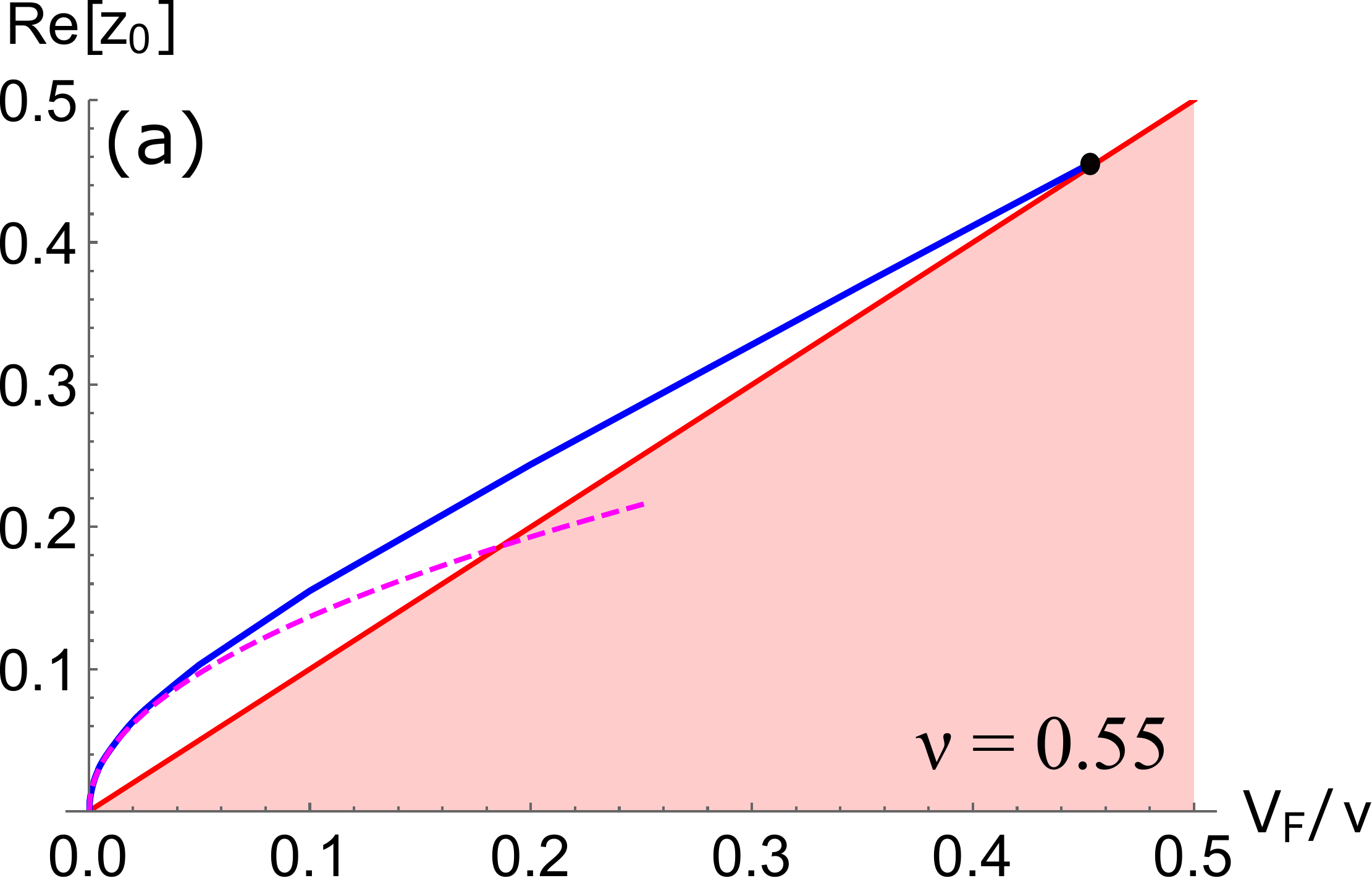} }}\\%
    \subfloat{{\includegraphics[width=0.87\linewidth]{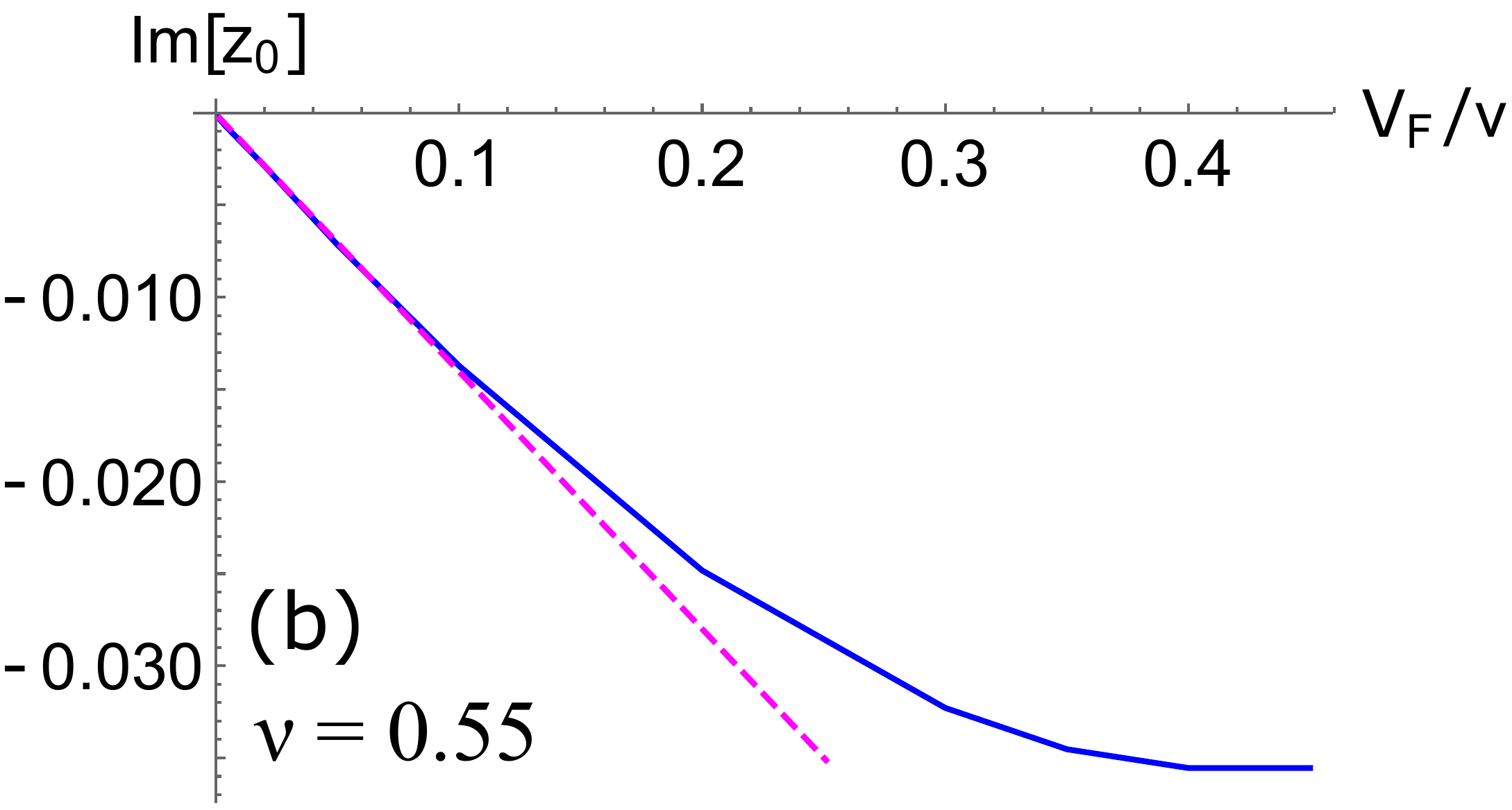} }}%
    \caption{ Numerical solution of the complex pole of the acoustic plasmon, $z_0$ vs. the velocity ratio, $\d_v = V_F / v$, obtained from numerically solving Eq.~(\ref{acoustic plasmon}) and using $\nu = 0.55$ [which corresponds to realistic parameters in Table~\ref{tab:parameters}]. (a) The real part of the solution, which gives the velocity of the acoustic plasmon in units of $v$. The dashed magenta line is the the approximate expression (\ref{acoustic plasmon_ana}). The red shaded region denotes the on set of the p-h- continuum of the holes. As can be seen for a constant $\nu$ increasing $V_F / v$ causes the pole to run into the p-h continuum of the holes where the solution is lost (marked by the black dot). (b) The imaginary part corresponding to the damping rate of the acoustic plasmon. We find that this mode is always weakly damped. The magenta dashed line is the approximate expression (\ref{acoustic plasmon_ana}).  }%
    \label{fig:ap}%
\end{figure}

By solve Eq.~(\ref{acoustic plasmon}) numerically we find that there exists a solution of the form (\ref{solution}), however, not for any $\d_v<1$. Namely, for a given value of $\nu$ there is a $\d_v$, above which, the physical solution of equation (\ref{acoustic plasmon}) disappears. For example, in Fig.~\ref{fig:ap} we plot the numerical solution of Eq.~(\ref{acoustic plasmon}) using the realistic values from Table \ref{tab:parameters} (giving $\nu \approx 0.55$). We find that the solution exists only for $\d_v < 0.45$.
Since the realistic parameters in Table \ref{tab:parameters} correspond to $\d_v \approx 0.4$, the acoustic plasmon pole exists within our isotropic approximation, but is close to the critical value for its disappearance. We also find that it is only weakly damped in the whole range where it exists.

To gain more intuition on this mode we can also solve for the acoustic plasmon in the limit of $\nu \d_v \rightarrow 0$, where the solution gives the known result~\cite{bennacer1989acoustic}
\be
u  =  \sqrt{ \nu  v V_F\over 3 } -i{\pi \nu V_F\over 12 }    \label{acoustic plasmon_ana}
\ee
This corresponds to the limit of $V_F / v \rightarrow 0$ in Fig.~\ref{fig:ap}, where the real part goes to zero like $\sqrt{V_F}$, while the imaginary part goes to zero linearly.
We can also obtain the effective electron coupling to the acoustic mode in this limit by expanding the interaction (\ref{Coulomb}) in the vicinity of the mode
\be
V(\w,q) \approx {{4\pi e^2 \over \ve_\infty N q_{TF}^2} \left[ 1 - {\left({u_1 q} \right)^2 \over (uq)^2 - \w^2}  \right] }\label{ac_pl_pole}
\ee
As noted earlier, this gives the same result as in the exchange of acoustic phonons in the jellium model, where the dimensionless coupling constant is given by $\lambda = \rho V(\w = 0) \approx 1/2N$.

It is important to note that in what follows we do not use the approximate form Eq.~(\ref{ac_pl_pole}) to estimate $T_c$, or in any other place in our calculations. We use the full RPA vertex Eqs.~(\ref{Coulomb}, \ref{epsilon}) which is dealt with numerically.

\section{Superconductivity}
We now turn to discuss the possible pairing instabilities due the collective modes of the electronic fluid in bismuth.
The frequency of these modes is higher than the corresponding Fermi energy of the holes and therefore we focus on superconducting
instabilities driven by the electron pockets, which have a higher Fermi energy. Thus the important role of the hole band in this model comes from its contribution to the RPA polarization (\ref{epsilon}).

To investigate the instability due to the interaction  (\ref{Coulomb}) we utilize the linearized Eliashberg equation (For a detailed derivation see Appendix~\ref{app:Eliashberg})
\be
\hat \D(i\w,\bs k) = -{T_c \over   L^3}\sum_{\w',\bs k'} {M_{\bs k,\bs k'}\hat \D(i\w',\bs k')M_{\bs k',\bs k} \over {\w'}^2 + \left[\e(k') \right]^2 } V(i\w -i\w',\bs k'-\bs k) \label{eliashberg}
\ee
where $\hat \D(i\w,k)$ is a $2\times 2$ matrix representing the order parameter in the two-dimensional basis of occupied bands.

The main difference between standard Eliashberg theory and Eq.~(\ref{eliashberg}) is the appearance of the rotation matrices $M_{\bs k,\bs k'}$, which project into the band basis of the two occupied bands, and thus adds non-trivial momentum form factors due to spin-orbit coupling.
For simplicity, here, we assume $s$-wave pairing, i.e. $\hat \D(i\w,\bs k)=\D(i\w,k)\,\md{1}$. In this case, tracing over both sides of Eq.~(\ref{eliashberg}) the gap equation assumes the form
\begin{align}
& \D(i\w, k) =\label{eliashberg2}\\& -{T_c \over  2 L^3}\sum_{\w',\bs k'}{\mrm{Tr}\left[ M_{\bs k,\bs k'}M_{\bs k',\bs k} \right] \D(i\w',k') \over {\w'}^2 + \left[\e(k') \right]^2 } V(i\w -i\w',\bs k'-\bs k)\nn
\end{align}
where
\begin{align}
 {1\over 2}\mrm{Tr}\left[ M_{\bs k,\bs k'}M_{\bs k',\bs k} \right]= {1\over 2} \left(1 +{v^2 \bs k \cdot\bs k'+\D_{bg}^2 \over \e_{ k}\e_{k'}} \right)\label{form factor}
\end{align}
This additional form factor is a consequence of spin-orbit coupling in the Dirac bands and comes from the transformation of the density operator to the Bloch-band basis (see Appendix ~\ref{app:Eliashberg}).

We also note that the form factor \eqref{form factor} can potentially reduce $T_c$ in the $s$-wave channel without breaking time-reversal symmetry or modifying the density of states at the Fermi level. For more details on this see Appendix \ref{andersons}.

\subsection{$s$-wave superconductivity from the dynamically screened Coulomb interaction }
Let us now turn to the main focus of the current work and consider the specific case of the screened Coulomb interaction (\ref{Coulomb}). As mentioned we consider, for simplicity, $s$-wave pairing, in which case the linearized Eliashberg equation (\ref{eliashberg}) reduces to the form
\be
 \Phi(i\w, k) = \sum_{\w'}\int d k' \mc K_{k,k'}^{\w,\w'} {\Phi(i\w', k')}\,.\label{eliashberg2}
\ee
where $\Phi(i\w,\bs k) = \D(i\w,\bs k)/ \D(0,k_F)$
The value of $T_c$ is obtained by finding an eigenvector of the kernel
\be
\mc K_{k,k'}^{\w,\w'} \equiv -{ \rho T_c \over  \e_F } {V_s(i\w-i\w',k,k')\over  \left({\w'/ \e_F}\right)^2 + {\left(\e_{k'}/\e_F -1 \right)^2} } \label{K}
\ee
which has unity eigenvalue. Here $\rho = k_F^2 / 2\pi^2v$ is the density of states per spin and pocket and
\begin{align}
V_s (i\w-& i\w',k,k') \equiv \label{avg_int}\\ &{1\over 2 }\int_{-1}^1 ds \,{1+ s \over 2}\,V\left(i\w-i\w',\sqrt{k^2 + k'^2 -2 k k' s}\right)\nn
\end{align}
is the interaction averaged over the solid angle between $\bs k$ and $\bs k'$ [including the form form factor (\ref{form factor}) taken in the limit $v k_F \gg \D_{bg}$].

\begin{figure}
 \begin{center}
    \includegraphics[width=1\linewidth]{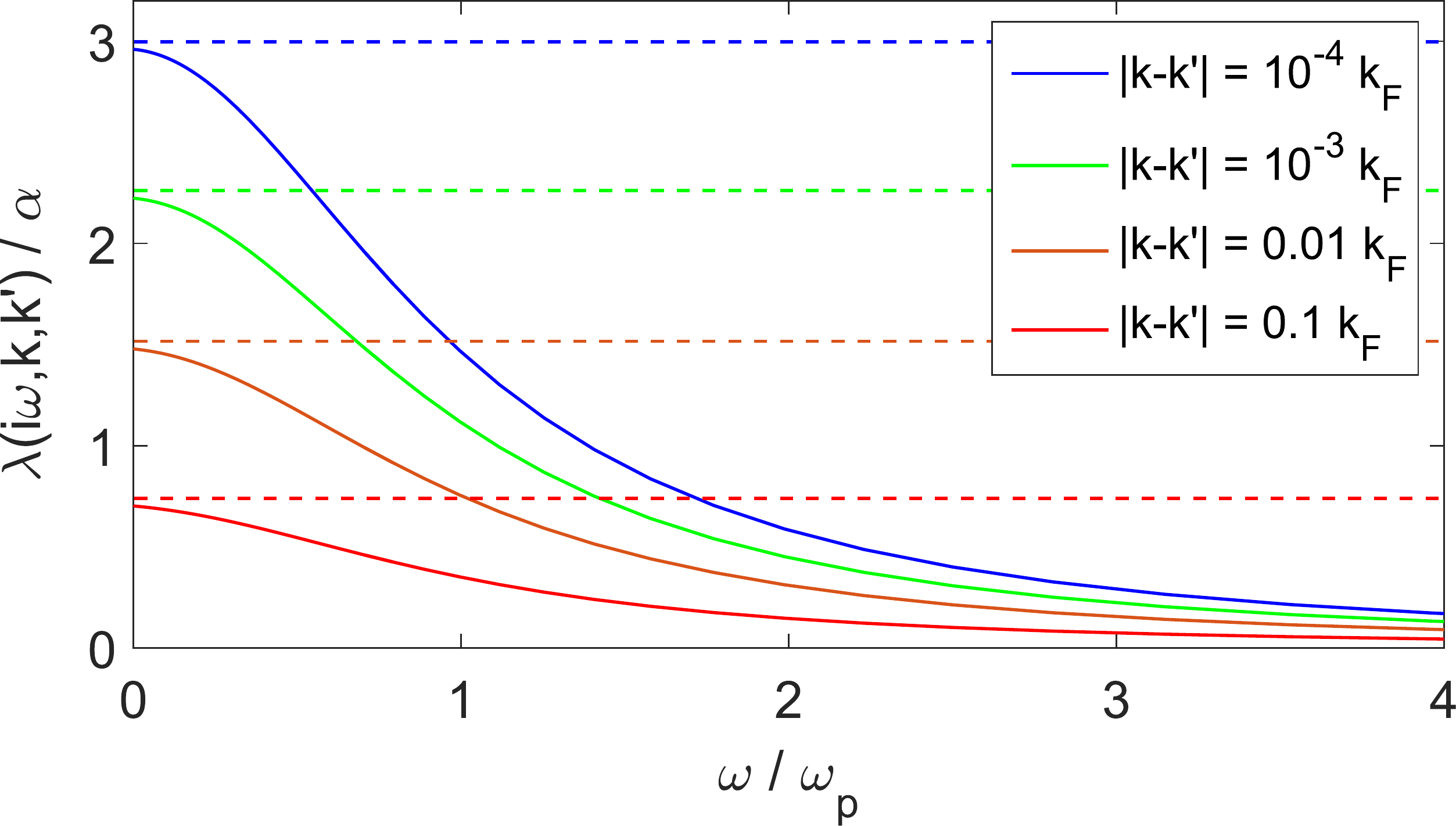}
 \end{center}
\caption{The attractive part (\ref{lam}) of the interaction (\ref{avg_int}) normalized by $\a$ vs. frequency in units of the plasma frequency, for different values of $|k-k'|$ near the Fermi energy and for the parameters in Table~\ref{tab:parameters}. The dashed lines represent the instantaneous repulsive part (\ref{mu}) for comparison and follow the same color coding.  }
 \label{fig:int}
\end{figure}

To make a comparison with standard Eliashberg theory (see for example Ref.~\cite{Margine2013}) we artificially decompose the interaction into two parts (note that we perform the calculation with the full vertex function \eqref{avg_int}, and this decomposition is only for the sake of discussion)
$
\rho V_s(i\w,k,k')= \mu(k,k') - \lambda(i\w,k,k')\,.
$
where
\be \mu(k,k') = \lim_{\w \rightarrow \infty}\rho V_s(i\w,k,k')\label{mu}
\ee
is the instantaneous Coulomb repulsion and
\be
 \lambda (i\w,k,k')= \rho V_s(i\w,\w',k) - \mu(k,k')
 \label{lam}
 \ee
is the retarded attractive interaction coming from the collective modes of the electron gas.
In Fig.~\ref{fig:int} we plot the attractive part (solid lines) of the interaction (\ref{lam}) normalized by the fine structure constant $\a$ for different values of $|k-k'|$ (using the parameters in Table~\ref{tab:parameters}. The repulsive part (\ref{mu}), corresponding to the same values of $|k-k'|$, and also normalized by $\a$ is plotted for comparison (dashed lines with corresponding colors). Here the integration over solid angle in Eq.~(\ref{avg_int}) is performed numerically.

It is important to note that the total potential \eqref{avg_int} is repulsive for {\it any} Matsubara frequency $\w$. This is the same in the conventional electron-phonon coupling, where it is crucial to renormalize the high-frequency repulsion, $\mu$, to a much smaller $\mu^*$, such that an effective attraction is obtained.

Let us now compare the interaction terms Eqs. (\ref{mu}, \ref{lam}), which originate from the electronic polarization (\ref{epsilon}), with the case of phonon mediated interactions.
First, we note that the interaction diverges logarithmically as $k$ approaches $k'$, as opposed to the phonon case where it is typically a constant. Thus, despite the weak coupling constant ($\a <1$) the attractive term can reach reasonably high values. This is crucial for superconductivity in the low density limit.

Second, the difference between the repulsion, $\mu(k,k')$, and the zero frequency limit of the attraction, $\lambda(0,k,k')$, is weakly dependent on $|k-k'|$ and is very small (it is given by the angular average (\ref{avg_int}) over the statically screened Coulomb interaction).
This implies that at small $|k-k'|$ the strength of the retarded attraction coming from $\lambda(i\w,k,k')$ is almost as strong as the instantaneous repulsion $\mu(k,k')$. In this scenario, even a small renormalization of the instantaneous repulsion at frequencies higher than the plasma frequency is sufficient to generate attraction.
On the other hand, as opposed to phonon superconductivity, the plasma frequency, $\w_p$, is of the same order as the Fermi energy, $\e_F$. Thus there is a very small window of energies between $\e_F$ and $\w_p$ where this renormalization may occur.

It is also important to note that in the weak coupling limit the form form factor (\ref{form factor}) suppresses the repulsion (\ref{mu}) more than it does to the attraction (\ref{lam}), and thus contributes to superconductivity. This is because the attractive part is mainly coming form the plasmon pole which exhibits a $1/q^2$ divergence. It is dominated by small $q$ scattering (i.e. $\bs k \simeq \bs k'$) and, in that limit the matrix element in Eq.~(\ref{form factor}) becomes unity and does not suppress the coupling constant.  On the other hand the screened repulsion coming from larger $q$ gets suppressed due to the angular dependence of (\ref{form factor}).

\subsection{Details of the numerical solution}
We solve Eq.~(\ref{eliashberg2}) numerically. First we discretize the momentum integral into $2W_k$ points $ k \in \{k_{-W_k} ,\ldots , k_{W_k}\} $, which are distributed around the Fermi momentum. $k_{W_k} = k_F+\L$ is the high momentum cutoff and $\L$ is taken to be $\L = 2k_F$.
We note that the value obtained for $T_c$ is very weakly affected by the value of the momentum cutoff which can also be set to be equal to or smaller than $k_F$.
What is however a crucial parameter is the density of points near the Fermi surface. We control the density of points using a power law distribution of points around $k = k_F$
\[|k_i - k_F|^{{1\over 1+\b}} - |k_{i-1} - k_F|^{{1\over 1+\b}} = {\L^{{1\over 1+\b}} \over W_k}\]
 where $\beta$ is the exponent defining the divergence of the density of points near the $k_F$ ($\beta = 0$ corresponds to a uniform distribution):
 \[p_k = {|k-k_F|^{-\beta} \over  (1+\beta)}\,.\]
 Here $p_k$ is the density of points.
Similarly, we define a set of Matsubara frequencies $\w \in \{\w_0 , \ldots, \w_{W_\w} \}$, where $\w_0 = \pi T$ is the lowest Matsubara frequency and the rest of the points are distributed according to
\[\w_i ^{1\over 1+\g} - \w_{i-1}^{1\over 1+\g} ={ (D\,\e_F)^{1\over 1+\g} \over  W_\w} \]
where $\g = 1/3$ is the divergence exponent for the divergence of Matsubara frequencies around zero $p_\w = \w^{-\g}/(1+\g)$ and $D$ defines the cutoff in units of the fermionic Fermi energy.

To obtain the transition temperature we reshape the kernel in Eq.~(\ref{K}) into a $(2W_k W_\w \times 2W_k W_\w)$ and solve for the eigensystem of this matrix~\cite{Takada1992}. $T_c$ corresponds to the point where the largest {\it positive} eigenvalue reaches unity.

\begin{figure}
 \begin{center}
    \includegraphics[width=0.9\linewidth]{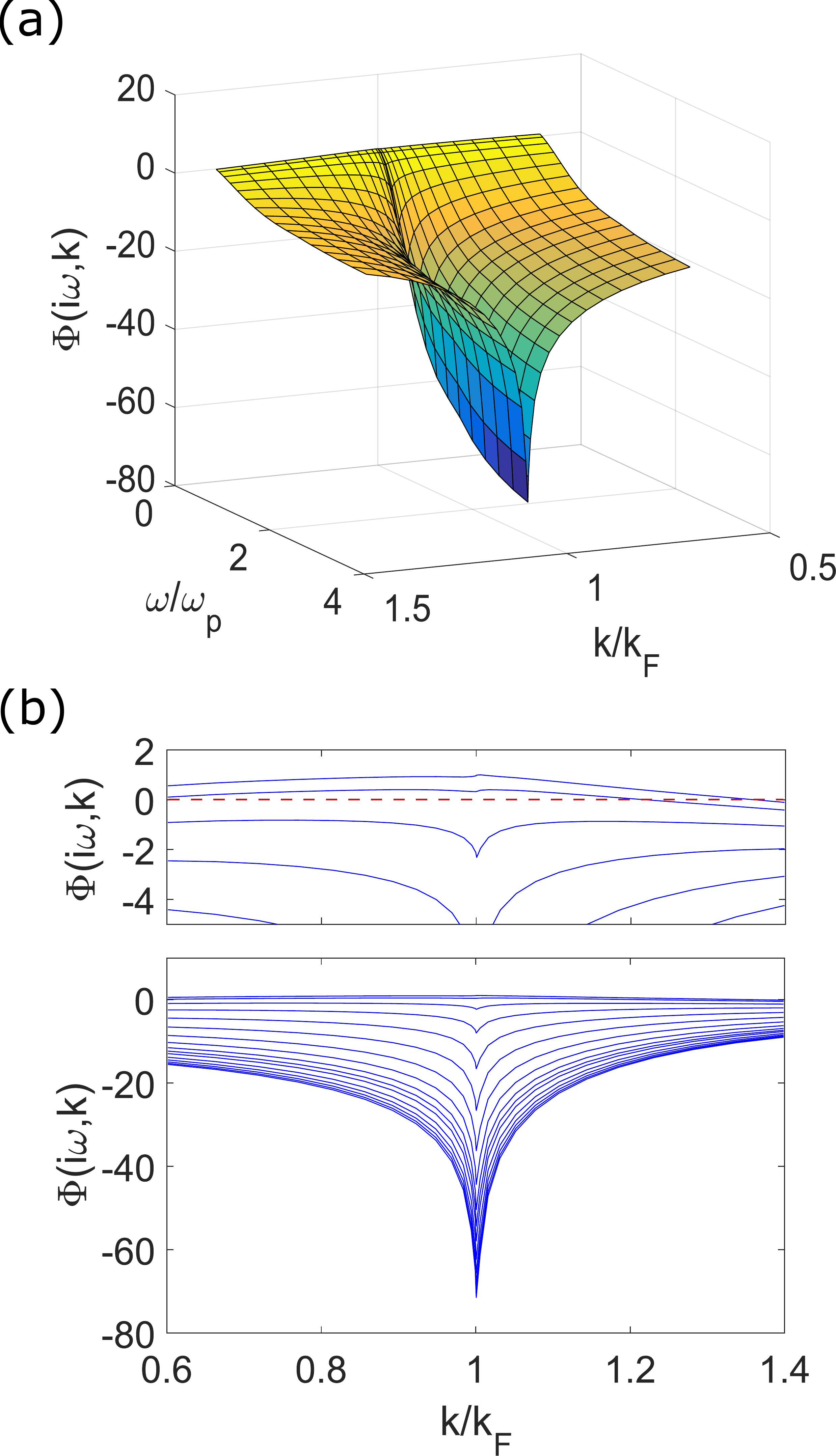}
 \end{center}
\caption{ (a) The eigenvector solution of Eq.~(\ref{eliashberg2}) as a function of momentum and frequency for $\beta = 2$, $\g = 0.5$, $W_k = 20$, $W_\w = 16$, $\L = k_F/2$, $D = 5$, $T = 300 \,\mrm{mK}$ and $\a = 1$. (b) Bottom: The same solution as a function of momentum for different frequencies (increasing from top to bottom; the lowest curve is $\w = 4 \e_F$). Top: Closeup on the lowest frequencies, exhibiting a peak near $k = k_F$. The dashed red line marks zero, showing that the gap at $\w = 0$ is positive and changes sign as the frequency is increased. Also note that it may change sign as a function of momentum.}
 \label{fig:gap}
\end{figure}
In Fig.[\ref{fig:gap}.a] we plot a typical eigenvector obtained from diagonlization of the kernel (\ref{K}) as a function of frequency and momentum, using $\beta = 2$, $\g = 0.5$, $W_k = 20$, $W_\w = 16$, $\L = k_F/2$, $D = 5$, $T = 300\,\mrm{ mK}$ and $\a = 1$.
In the bottom part of Fig.[\ref{fig:gap}.b] we plot the same eigenvector vs. momentum for different frequencies.
At high frequency the gap function is negative and changes sign as the frequency is reduced, consistent with Anderson-Morel~\cite{Morel1962} type picture. However, as pointed out by Takada (e.g. \cite{Takada1992}), in this case the finite-frequency negative part of the solution, is sharply peaked at $k = k_F$ and is much larger than the positive part at the lowest Matsubara frequency $\w = \pi T$. In the top panel of of Fig.[\ref{fig:gap}.b] we show a closeup of the bottom panel showing that at small frequency the eigenvector becomes positive.

\subsection{The transition temperature}
\begin{figure}
 \begin{center}
    \includegraphics[width=1\linewidth]{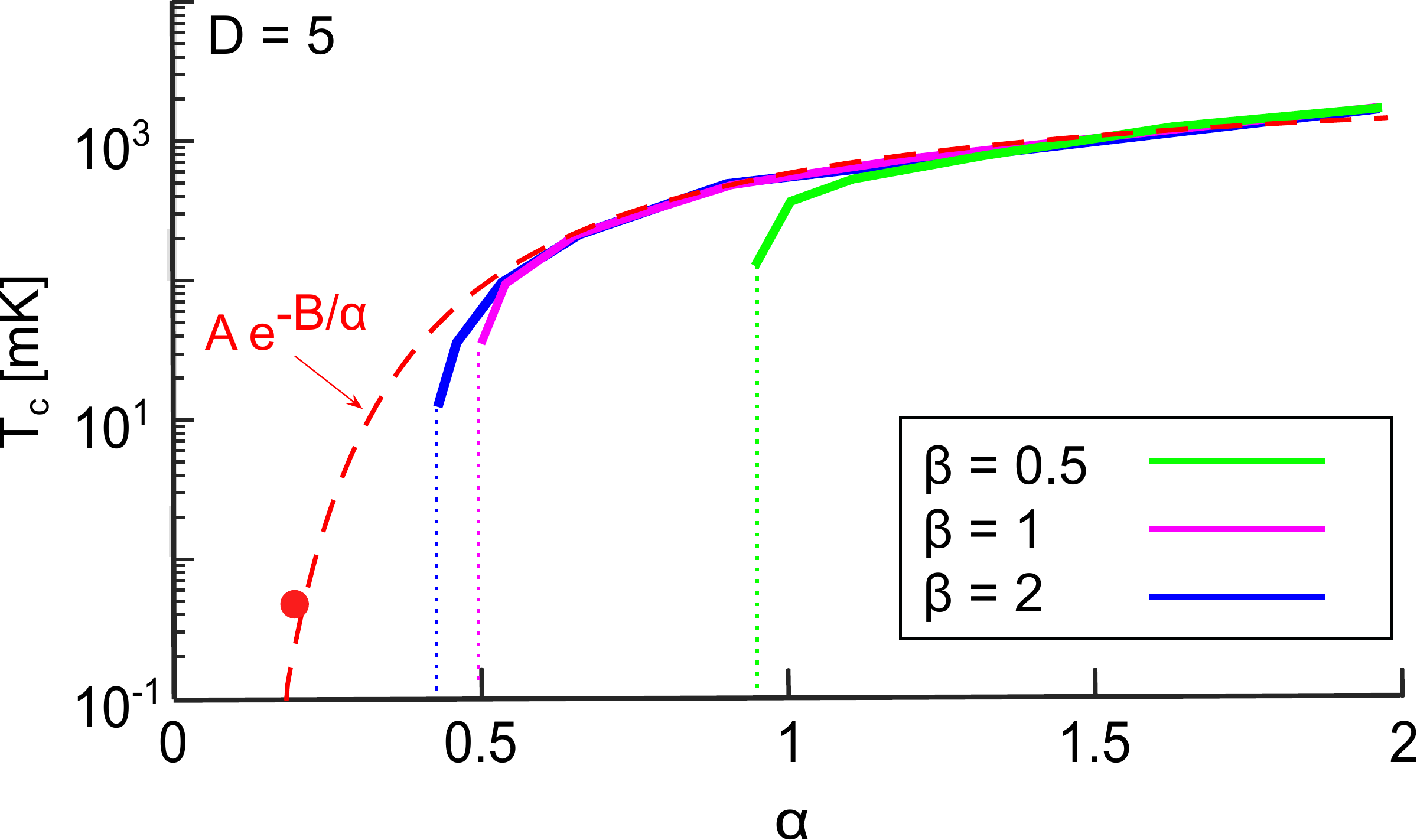}
 \end{center}
\caption{The transition temperature $T_c$ as a function of the fine structure constant $\a$ for three different density of points $\b = 0.5,1\,\mrm{and}\,2$ (corresponding to deep blue, light blue and cyan, respectively) and using $D = 5$, $W_k = 14$, $W_\w = 16$ and $\g = 0.5$. The red (dashed) line is a fit using $T_c =  e^{-B/\a}$, with $A = 4K$ and $B = 1.9$ for comparison. The yellow corresponds to $\a = 0.2$ and $T_c = 0.5\,\mrm{mK}$. Plugging $\a = 0.2$ in Eq.~(\ref{wp}) gives $\w_p = 18\mrm{meV}$, which is very close to the experimental value ~\cite{Tediosi2007}. }
 \label{fig:Tc}
\end{figure}

In Fig.~\ref{fig:Tc} we plot the transition temperature, $T_c$, vs. the fine-structure constant, $\a$, for three different values of $\b=0.5,1$ and $2$. For comparison, the smallest distance between points corresponding to these values of $\beta$ is $\min |k_i/k_F - 1| = 0.013,0.0025 $ and $0.0001$, respectively.
To generate this plot we used $W_\w = 16$, $W_k = 14$, $\g = 0.5$, $\L = k_F$ and $D = 5$.

We find that at higher values of $\a$, $T_c$ roughly follows an exponential form $T_c = B\, e^{-A/\a}$ (The exponential form is plotted for comparison with $A = 1.9 $ and $B = 4 K$, see dashed line). However, at lower values of $\a$, $T_c$ sharply drops to zero at a value which depends on the density of points near $k = k_F$, $\b$.
Naively, this implies that there is a minimal $\a$ for superconductivity.
However, since increasing the density of points near $k = k_F$, $\b$, which is not a physical parameter, enhances the regime where the exponential decay is observed, we argue that it is also possible that the superconducting instability exists for any small $\a$. Numerically, however, we can not obtain a solution at an arbitrarily low $\a$. Using the values of $A = 1.9 $ and $B = 4 K$, we find that the measured $T_c = 0.5 \, mK$ in bismuth is obtained for $\a = 0.2$, which is consistent with our earlier estimate [see discussion below (\ref{alpha})].

Here it is important to compare our results to the work of Takada~\cite{Takada1992}, which predicts a minimal $r_s$ for superconductivity despite integrating up to a very high cutoff, $D > 500$. The value of $\b = 0.5$ corresponds to the smallest density of points that Takada used in his work, which captures accurately the strong coupling limit but not the weak coupling one (as shown in Fig.~\ref{fig:Tc}). Therefore we argue that his minimal $r_s$ for superconductivity is an artifact of the density of points that he chose.
We also note as opposed to Ref.~\cite{Takada1992}, here there is also the heavy hole band, which as we shall see, greatly enhances $T_c$ in the weak coupling limit.

\begin{figure}
 \begin{center}
    \includegraphics[width=1\linewidth]{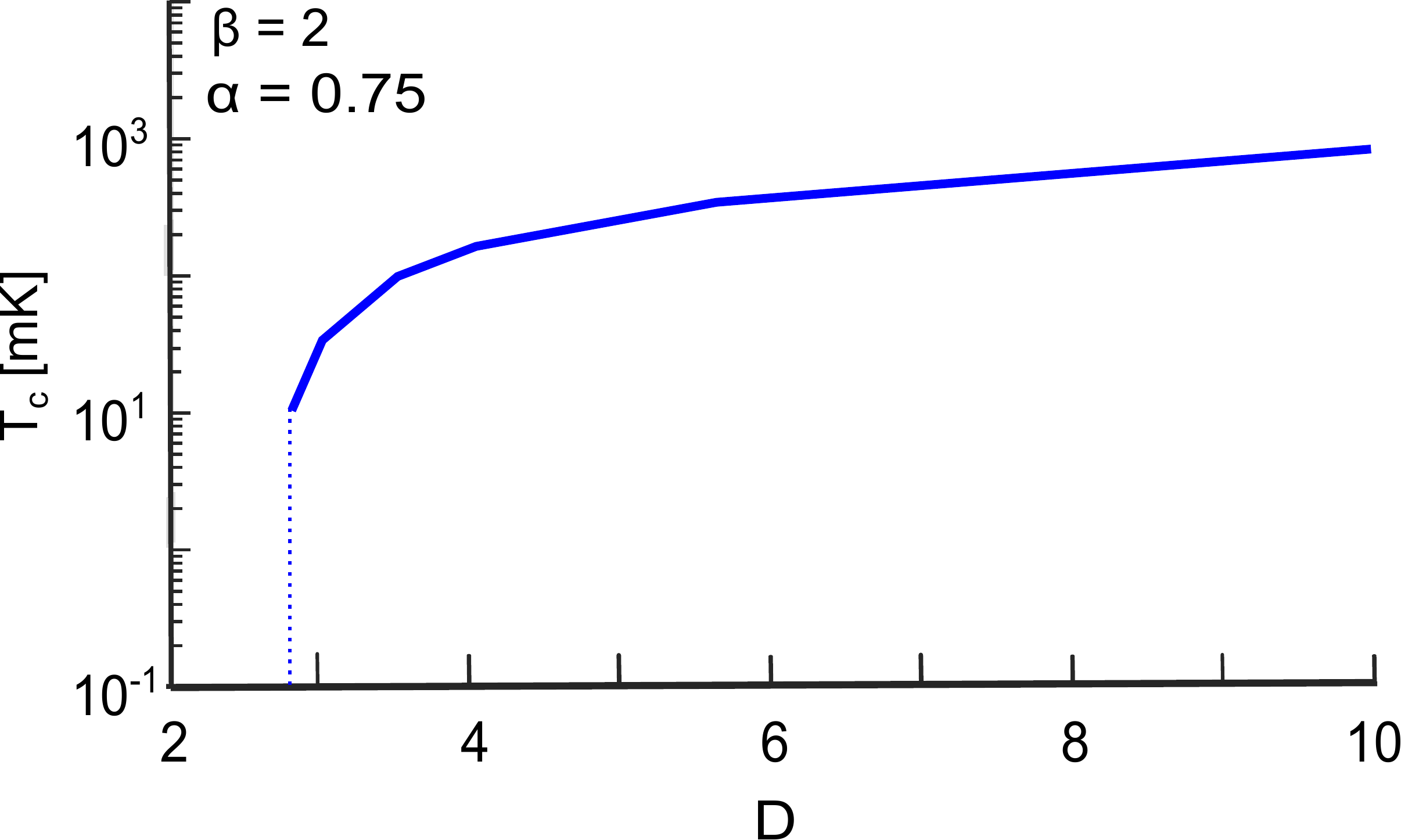}
 \end{center}
\caption{The transition temperature vs. the cutoff $D$. In this plot we used $\a  = 0.75$, $W_k = 14$, $W_\w = 16$, $\g = 0.5$ and $\b = 2$. }
 \label{fig:Tc_cutoff}
\end{figure}

Finally, we point out that here we have not used any phenomenological parameter such as the parameter describing the renormalization of the Coulomb repulsion in conventional Eliashberg theory, $\mu^*$~\cite{Margine2013}. However, we have used a high energy cutoff $\W = D\e_F$.
We find that we always need to use $D >1$ (i.e. to integrate to energies higher than $\e_F$) to find an instability. Since the processes taken into account in the Eliashberg theory (\ref{eliashberg2}) are not necessarily the most dominant contributions in this limit, $D$ should be regarded as an (implicit) phenomenological parameter, equivalent to $\mu^*$.
To get a better understanding of its effect on $T_c$ we plot the transition temperature as a function of $D$ in Fig.~\ref{fig:Tc_cutoff}. Here we have used $\a = 0.75$, $W_k = 14$, $W_\w = 16$, $\b = 2$ and $\g = 0.5$. As can be seen, increasing the cutoff $D$ enhances $T_c$ exponentially.

\section{Discussion}
\begin{figure}
 \begin{center}
    \includegraphics[width=1\linewidth]{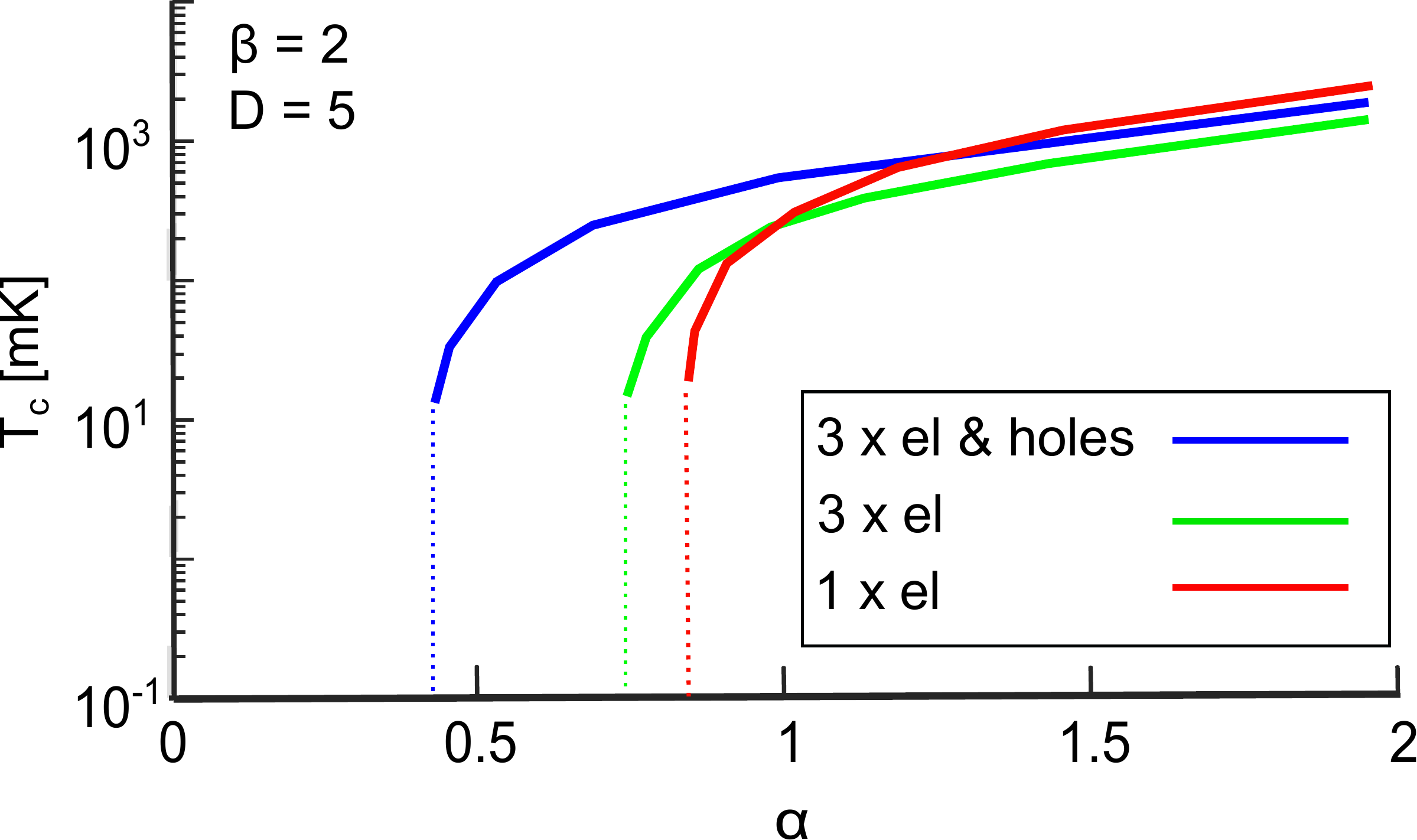}
 \end{center}
\caption{ The transition temperature $T_c$ vs. the fine structure coupling constant, $\a$ for three different cases: (i) with three electron pockets ($N = 3$) and the hole pocket, (ii) only the three electron pockets and (iii) just a single electron pocket ($N = 1$).   }
 \label{fig:Tc_noholes}
\end{figure}

We now turn to discuss the results. In the previous section we have presented evidence for a weak instability.
To understand the role of the multiple bands in bismuth on superconductivity we plot the transition temperature, $T_c$, vs. the fine-structure constant, $\a$, for three different cases (see Fig.~\ref{fig:Tc_noholes}):
(i) with three electron pockets and the single hole pocket, (ii) only three electron pockets and (iii) only, a single electron pocket. By comparing the three, we find that in the weak coupling limit the hole band enhances $T_c$, much more than the multiplicity of electron bands does. On the other hand, for high $\a$ the transition temperature reduces as we increase the number of bands.

\begin{figure}
 \begin{center}
    \includegraphics[width=1\linewidth]{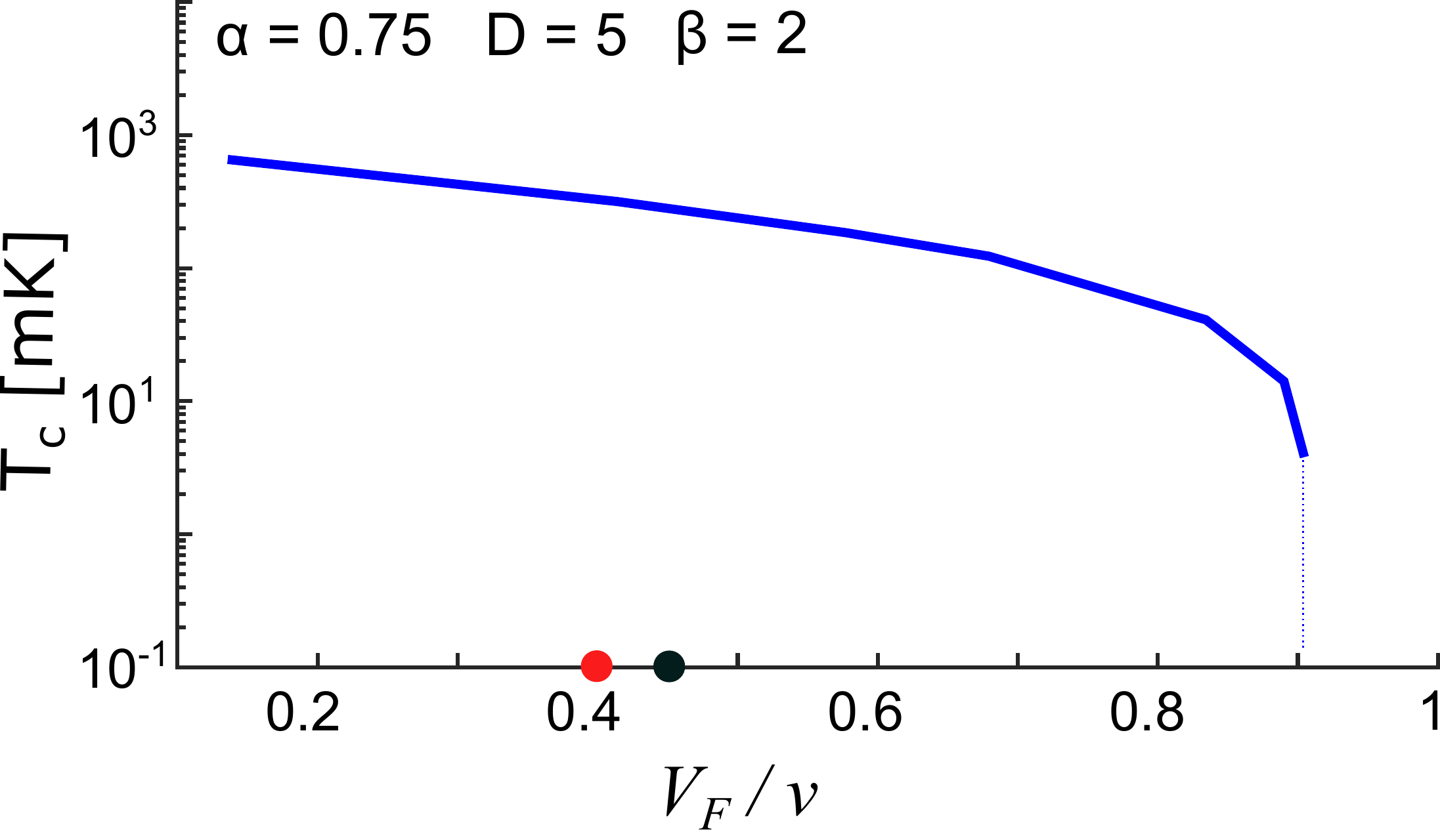}
 \end{center}
\caption{ The transition temperature, $T_c$, vs. the velocity ratio $\d_v = V_F/v$ for $\a = 0.75$, $\b = 2$ and $D = 5$. The red dot represents the value of $\d_v = 0.4$ which corresponds to the parameters in Table \ref{tab:parameters}. The black dot corresponds to the point where the acoustic plasmon pole disappears (see Fig.~\ref{fig:ap}. a).   }
 \label{fig:Tc_V_ratio}
\end{figure}

The strong influence of the holes at weak coupling may come from the existence of the additional acoustic plasmon mode~\cite{Garland,RADHAKRISHNAN1965247,pashitskii1968plasmon,frohlich1968superconductivity,ruvalds1981there,entin1984acoustic}. To investigate the role of the acoustic plasmon we calculate the dependence of $T_c$ on the velocity ratio $\d_ v  = V_F / v$ in Fig.~\ref{fig:Tc_V_ratio}. We find that the transition temperature decreases with increasing $\d_v$. However, as shown in Fig.~\ref{fig:ap}. (a) above a critical value of $\d_v = 0.45$, the solution of the acoustic plasmon no longer exists. Nonetheless, the dependence of the transition temperature on $\d_v$ is smooth, and no special feature is observed at that point. Therefore, we conclude that the acoustic plasmon is not the main driving force in the large enhancement of $T_c$ at weak coupling (shown Fig.~\ref{fig:Tc_noholes}).

\begin{figure}
 \begin{center}
    \includegraphics[width=1\linewidth]{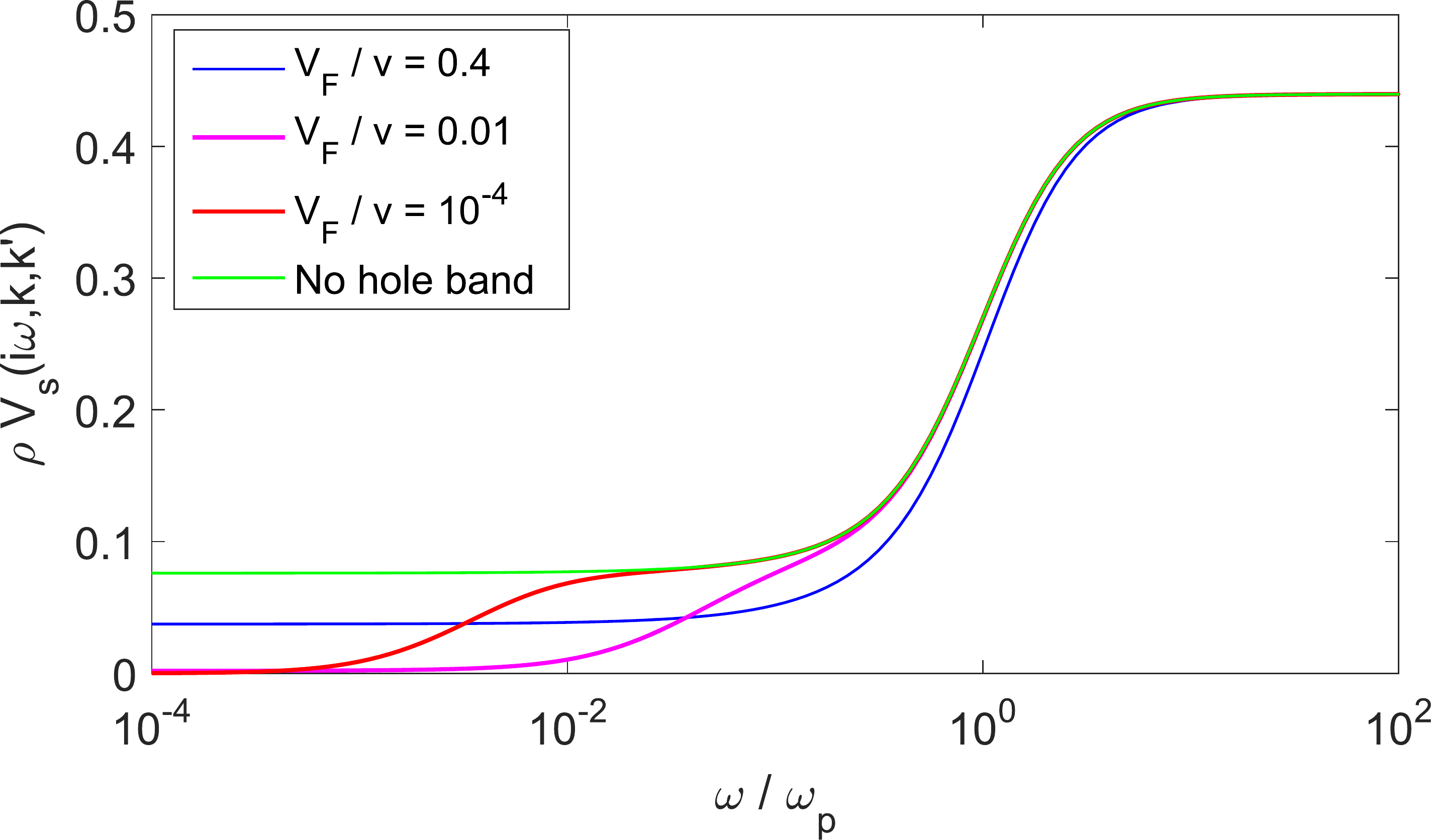}
 \end{center}
\caption{The averaged interaction (\ref{avg_int}) vs. frequency for three values of $\d_v = V_F/v = \,0.4$, $0.01$, $ 10^{-4}$ and the case where the hole band is absent. Here we took $|k-k'| = 0.25 k_F$. As can be seen $T_c$ goes down smoothly with increasing $\d_v$ around this point.  In the case of $\d_v = 0.01$ and $ 10^{-4}$ the retardation effect of the acoustic plasmon i clearly seen at $\w \sim0.1$ and $ 10^{-2}\,\w_p$, respectively.  }
 \label{fig:int2}
\end{figure}

The origin of this large enhancement becomes clear when considering the averaged interaction of (\ref{avg_int}) in the static limit. In Fig.~\ref{fig:int2} we plot the interaction (\ref{avg_int}) for $\d_v = 0.4$, $0.01$, $10^{-4}$ and for the case where there is no hole band. As can be seen the main difference between the curves is the saturation value in the static limit, which can be estimated to
\be
\rho V_s(0 ,k,k') = {q_{TF}^2 \over 8 k_F^2} \log \left[ \left(1- {\kappa_{TF}^2 \over 4 k_F^2}   \right)\log\left( 1+ { 4 k_F^2\over\kappa_{TF}^2 } \right)-1\right] \label{sat}
\ee
This value monotonically decreases with increasing $\kappa_{TF}$ [defined in Eq.~(\ref{kappa})]. Whereas the high frequency limit is the same in all cases. This implies that in all four cases the instantaneous repulsion (\ref{mu}) is the same. On the other hand, by definition, when the the saturation value at low frequency is larger it implies that (\ref{lam}) is smaller. Thus, in the case where there is no hole band the saturation value is higher because the contribution of the hole band to $\kappa_{TF}$ is vanishing, i.e. $Q_{TF} = 0$, which implies that the overall attractive interaction is weaker.

The reason the hole band is so much more effective in this enhancement than the electrons (see Fig.~\ref{fig:Tc_noholes}) is that at small $\d_v$ the holes greatly enhance $\kappa_{TF}$ while almost not modifying the total plasma frequency $\w_p \propto \sqrt{N + \d_k^2 \d_v}$ [see Eq.~(\ref{wp}, \ref{kappa})]. As we have already established, larger $\kappa_{TF}$, implies greater attraction. However, it is also crucial that the hole band does not enhance the plasma frequency. The reason is that when the plasma frequency is enhanced it reduces the range of frequencies between the cutoff $\W = D \e_F$ and $\w_p$ where the repulsive part (\ref{mu}) can be renormalized.

As mentioned earlier, in the limit $\d_v \ll 1$ the acoustic plasmon becomes the acoustic phonon in the "jellium" model. We can see clearly the retardation effect below the scale of the acoustic plasmon frequency. Note that for the realistic value of velocity ratio, $\d_v = 0.4$, no such separation of scales is visible.

\subsection{Experimental consequences}
So far we have argued that the attractive interaction in bismuth is generated only from the dynamically screened Coulomb interaction. In this section we discuss the experimental consequences of this prediction.

First, we note that the mass of the bismuth atom did not play any role in our theory. Therefore, the observation of any isotope effect will falsify our theory. It is important to note that measuring the isotope effect in bismuth is challenging due to the large atomic mass. The four stable isotopes of bismuth are Bi$^{207}$, Bi$^{208}$, Bi$^{209}$ and Bi$^{210\mrm m}$, which allows a variation of more than $1\,\%$ in the mass. The accuracy of the measurements is close to this value and hopefully a conclusive measurement can be done.

Another possible effect is the appearance of a resonant spectral feature in the tunneling density of states above the superconducting gap~\cite{Schrieffer1963}.
This is the fingerprint of a resonant plasmon excited by the tunneling electron, and will therefore appear at the plasma frequency~\cite{Ruhman2016} and is expected to change when going through the transition. Such a feature has been observed in optical infra-red reflectivity measurements performed at higher temperatures~\cite{Tediosi2007}, signaling the strong electron-plasmon coupling in bismuth.

Finally, according to our predictions the density of states of the holes enhances $T_c$. It would be interesting to see whether the application of uniaxial strain, which enhances the hole mass, can enhance $T_c$.

\section{Conclusions}
We have studied superconductivity in Bismuth. We argued that at low carrier concentration only long-ranged interactions are capable
of causing such an instability. In the absence of any experimental evidence of a critical point we investigated the more likely scenario in which the dynamically screened Coulomb repulsion gives rise to an effective retarded attraction on the energy scale of the longitudinal plasma oscillations.
We used an approximate isotropic band structure and the random phase approximation for the screened Coulomb interaction. Within these approximations we found that there is a weak coupling instability.

The transition temperature is greatly enhanced by the existence of a heavy hole band. We showed that this enhancement is due to the large mass of the holes, which allows for an enhancement of the static screening (Thomas-Fermi) without enhancing the plasma frequency. We showed that $T_c$ is not dramatically decreased when the acoustic plasma mode is absent. Therefore, we concluded that it is not the main contributor to attractive interactions in bismuth.

Interestingly, we found that the superconducting instability found by Takada~\cite{Takada1992} can be extended to lower coupling strength by enhancing the density of mesh points near the Fermi surface. However, we still observed a minimal coupling strength for the instability. Thus, an interesting question that remains open, but is much motivated by our study, is whether an instability exists at arbitrarily weak coupling?

In this work we have focused on $s$-wave superconductivity. However, as pointed out by Refs.~\cite{Takada1992,takada1993s,Lederer2015,Metlitski2015} when the attractive interaction is long ranged (small $q$ scattering) the coupling strength in the higher angular momentum channels is comparable to the $s$-wave channel. In this scenario the symmetry of the order parameter is mainly dictated by the short ranged interactions which are not captured by Eq. (\ref{Coulomb}). Therefore, we conclude that one can not rule pairing of higher angular momentum channels in bismuth.

Another interesting product of our work is that we find that when rotating the band basis to the Dirac bands and projecting to the occupied states an additional form form factor appears in the Eliashberg equation. This form factor aids superconductivity by suppressing the repulsion (\ref{mu}). Interestingly, we have also shown that this factor can affect $T_c$ in the $s$-wave channel without breaking time-reversal symmetry, depending on the ratio between the Dirac energy and the mass gap, $v k_F / \D_{bg}$.
Therefore, the case of a Dirac cone goes beyond what was considered by Anderson in his original theory~\cite{anderson1959theory} and opens the question, what is the effect of disorder in the Dirac mass, and other Dirac bilinears, on superconductivity in semimetals?

The same Anderson's theorem, concerning the effect of disorder on $T_c$, also does not hold in this case because of the strong $1/q^2$ dependance of the attractive interaction. The reason here is that disorder strongly effects small angle scattering, causing smearing of the scattering amplitude over a wider window of momenta. Thus, even the $s$-wave channel can be significantly affected by chemical potential disorder. Therefore, we point out that the influence of disorder on the transition temperature in plasmonic superconductors is not the same as in conventional superconductivity, and therefore should not be used to infer the symmetry of the gap without further study.

\section{Acknowledgments}
We thank Srinivasan Ramakrishnan, Lucile Savary, Jorn Venderbos, Inti Sodemann, Yuki Nagai, Brian Skinner and Anshul Kogar for many helpful discussions and for pointing out important papers. We also thank Nandini Trivedi and Andrew Dane for pointing out Ref.~\cite{Prakash2016}.
JR acknowledges a scholarship by the Gordon and Betty Moore Foundation under the EPiQS initiative under grant no. GBMF4303. PAL acknowledges the support of DOE  under grant no. FG02-03ER46076.


\begin{appendices}

\section{Electron and hole polarizations} \label{app:polarizations}
In this section we write the explicit formula for the electron and hole polarization appearing in the dielectric function Eq. (\ref{epsilon}). These polarizations are calculated from the standard formulae
\begin{widetext}
\begin{align}
\Pi_e(i\w,q) &= -{1\over 2\Omega}\sum_{\bs k}  {n_F(\e_k )-n_F(\e_{k+q} )\over -i \w +\e_k -\e_{k+q}}\left(1+ {\e_k^2 + v ^2 \bs k \cdot \bs q \over \e_k \e_{k+q}} \right)\\
\Pi_h(i\w,q) &= -{1\over \Omega}\sum_{\bs k}  {n_F(E_k )-n_F(E_{k+q} )\over -i \w +E_k -E_{k+q}}
\end{align}
Note the additional Berry-curvature form factors in $\Pi_e$.
We note that the form factor appearing in the first equations is the same as (\ref{form factor}).
Neglecting the small band mass, $\D_{bg}$, of the electron pockets, the exact formula for these polarization bubbles is given by
\begin{align}\label{e_polarization}
\Pi_e (i\w,q) =2N\rho \left\{ \begin{array}{@{}ll@{}ll@{}}
    \displaystyle {  {1\over 12}\left[ {3 \tilde \w^2+ 8 \tilde q^2 -36 \over 3} -{\tilde \w \over \tilde q} \left( \tilde \w^2 +3\tilde q^2 -12 \right) \arctan{\tilde q \over \tilde \w} \right] } & ;& \;\text{if}\ \tilde q < 1\\ && \\
    \displaystyle {1\over 72 \tilde q} \left[6\tilde \w^2 +12 \tilde q \left( \tilde q- 2 \right) -44 +3\tilde \w \left( \tilde \w^2 +3\tilde q^2 -12\right) \left\{ \tan^{-1}{\tilde q -2 \over \tilde \w }-  \tan^{-1}{\tilde q  \over \tilde \w }\right\} \right.  & ;&\;\text{otherwise}  \\
    \displaystyle{+\left. 6\left( 4+\left(\tilde q-3 \right)\tilde q^2 -3\tilde\w^2 \right)\tan^{-1}\left({2-2\tilde q \over 2+\tilde q \left( \tilde q-2\right) +\tilde \w^2} \right) \right] }&  &\;
  \end{array}  \right.
\end{align}
and
\be
\Pi_h (i\w,q) = {2R\over 16}\left[ 8 - {4\tilde Q^2 -\tilde Q^4 + \tilde W^2 \over \tilde Q^3} \log \left(1- {8 \tilde Q^3 \over \tilde Q^2 \left( 2+\tilde Q \right)^2+\tilde W ^2} \right) + 2i {\tilde W \over \tilde Q} \log\left({\tilde W^2-4\tilde Q^2 +\tilde Q^4 + 4i \tilde Q \tilde W\over \tilde W^2-4\tilde Q^2 +\tilde Q^4 - 4i \tilde Q \tilde W} \right) \right] \label{h_polarization}
\ee

\end{widetext}
where $\rho ={ k_F^2 / 2\pi^2 v}$, $R ={K_F^2 / 2\pi^2 V_F}$ are the density of states per spin and pocket, $\tilde \w = \w / \e_F$, $\tilde q = q/k_F $, $\tilde W = \w/E_F$ and $\tilde Q = q/ K_F$.

\section{Collective modes in a two fluid model}\label{app:collective modes}
In this section we derive the effective interaction in the vicinity of the collective modes of the electron-hole plasma. In this case, where there are two fluids with significantly different Fermi velocities, there are two fermi-surface volume modes: the gapped plasmon and the acoustic plasmon. These modes are obtained by seeking the zeros of the dielectric function (\ref{epsilon}), i.e. by solving the equation
\be
q^2   = N q_{TF}^2 \Pi_e(\w,q) + Q_{TF}^2\Pi_h (\w,q) \,.\label{poles}
\ee
To get an intuitive picture, however, it will be sufficient to focus on the limit of $q \ll k_F$ where Eq.~(\ref{poles}) reduces to
\be
{q^2 \over N q_{TF}^2} = f(z)  + {\nu\over \d_v} f( z  / \d_v) \label{poles2}
\ee
where $\nu \equiv\d_k ^2 / N $, $z = \w / v q$, $\d_v = V_F / v$ and $f(z) = {z\over 2} \log \left({z+1 \over z-1} \right)-1$.

\subsubsection{The gapped plasmon}
The first mode, the gapped plasmon, occurs in the high frequency regime  $\w \gg v q ,\, V_F q$, where the electron and hole polarization are deep in the dynamic limit $\Pi_{e,h}(i\w,q) \propto q^2 / \w^2$. Therefore we find a solution of Eq.~(\ref{poles2}) at $\w = \w_p$ where $\w_p = \sqrt{w_p^2 + W_{p}^2}$, $w_p  = \sqrt{N/3} \, v\,q_{TF} $ and $W_p = \d_v\,\sqrt{ \nu }\,\w_p  $. The resulting interaction Eq. (\ref{Coulomb}) assumes the form
\be
V(\w,q)\approx {4\pi e^2 \over \ve q^2} \left[1 - {\w_p^2 \over\w_p^2 -\w^2}\right] \label{plasmon:app}
\ee
Plugging in realistic parameters for the averaged velocities and Thomas-Fermi momenta, $v$, $V_F$, $q_{TF}$ and $Q_{TF}$, we find that to match the experimental value of the plasma frequency, $\w_p = 18\, \mrm{meV}$, we need $\ve \approx 30$. (Note the significant deviation from the measured dielectric constant, $\ve = 90$~\cite{Boyle1960} which is due to the isotropic approximation).

Eq.~(\ref{plasmon:app}) is known as the {\it plasma pole approximation}, which captures the position of the plasma pole in frequency space but not the effective electron-plasmon coupling strength. The main reason for this inaccuracy is that at higher momentum, $q > q_c$, the plasmon runs into the p-h continuum of the electron bands and becomes strongly damped (see Fig.~\ref{fig:modes}). Denoting that $q_c\gtrsim \w_p / v$~\cite{Gurevich1962} we can obtain a better approximation for the plasmon coupling by limiting the  the phase space of
\be
V(i\w,q) \approx \left\{
  \begin{array}{@{}ll@{}ll@{}}
    \displaystyle {4\pi e^2  \over\ve_\infty\left( q^2 + \kappa_{TF}^2 \right)  } & ;& \;\text{if}\ q > \w_p / v\\
    \displaystyle {4\pi e^2  \over \ve_\infty q^2  } \left[ 1 - \a(q) {\w_p ^2 \over \w_p^2 - \w^2}\right] & ;&\; \text{otherwise}
  \end{array}\right. \label{interaction_approx}
\ee
where $\a(q) = \kappa_{TF}^2 / (q^2 + \kappa_{TF}^2)$ interpolates to the value of the screened Coulomb interaction in the static limit $\w \rightarrow 0$.

From Eq.~(\ref{interaction_approx}) we see that the plasmon response is observed in a window of scattering angles on the Fermi surface defined by $\cos \t_{\bs k,\bs k'} > 1-\w_p^2 /2 \e_F^2$.

\subsubsection{The acoustic plasmon}
The second mode, the acoustic plasmon,  is observed at lower frequencies $vq>\w>V_F q$. This limit lies within the p-h continuum of the electrons (see Fig.~\ref{fig:modes}), which means that the function $f(z)$ non-zero imaginary part $\mrm{Im}f(z) = \pi z/2$ and therefore we must seek a solution of Eq.~(\ref{poles2}) in the complex plane $z = z_0$ \cite{pines1956electron,bennacer1989acoustic}, such that
\be
v z_0 = u_1 + i u_2
\ee
and where $u_1$ and $u_2$ are real positive numbers and where $\d_v  <u_1 <1$.

In Fig.~\ref{fig:ap} we plot the numerical solution of the damped pole vs. values the velocity ratio $\d_v$.
The shaded region in panel (a) marks the onset of the particle hole continuum of the holes (where $\mrm{Im}f(z/\d_v) \ne 0 $). Thus, in this regime there is no longer a solution.
Note that using the parameters in Table \ref{tab:parameters} we get $ \nu = 0.5$ and $\d_v = 0.4$.

We find that the acoustic plasmon, for these parameters, is weakly damped. The velocity of this mode increases with $\d_v$. However, for $\d_v > 0.45$ the solution disappears and the only solution of Eq.~(\ref{poles2}) is the gapped plasmon.

We can also obtain an analytic solution of the acoustic plasmon in the limit of $\d_v \ll 1$. In this case Eq.~(\ref{acoustic plasmon}) assumes the form
\[z^2 \left(1-{i\pi z \over 2} \right) = {\d_v \nu \over 3 }\]
In the limit of small $\nu \d_v \ll 1 $ this gives the well known result\cite{bennacer1989acoustic}
\be
z = \sqrt{ \nu\d_v  \over 3} + i {\pi \nu \d_v \over 12}
\ee

\begin{widetext}
\section{Eliashberg theory}\label{app:Eliashberg}
\subsection{The action in Nambu space}
As a preparatory step for the Eliashberg theory we transform the Hamiltonian to an action in Nambu space.
Comparing the plasma frequency with the Fermi energy of the holes and electrons in bismuth (Table.~\ref{tab:parameters}) we find that the plasmon is only retarded with respect to the electrons. Therefore we will focus on the electron pockets here.

\subsubsection{Free part}
A single Dirac pocket, Eq.~(\ref{H_e}), is written in Nambu space as
\be
\mc S_0 = {1\over 2}\sum_{\bs k,\w} \Psi_{\bs k}^\dag \begin{pmatrix}
- i \w + H_e(\bs k)-\e_F & 0 \\
0 & - i\w -  H_e^* (-\bs k) +\e_F
\end{pmatrix}\Psi_{\bs k}\label{S0}
\ee
where the Nambu spinor are given by
\[\Psi_{\bs k} =\begin{pmatrix}\psi_{\bs k} \\ \psi_{-\bs k}^\dag \end{pmatrix}\,.\]
and $\psi_{\bs k} = \left( \psi_{\bs k,1},\psi_{\bs k,2},\psi_{\bs k,3},\psi_{ \bs k,4}\right)^{T}$ is written in the notations of Re.~\cite{Wolff1964}.
The Hamiltonian is time-reversal symmetric, and therefore $\mc T  H^*(-\bs k) \mc T ^{-1} = H(\bs k)$,  where $\mc T = i \, \s^y \, s^z$. Therefore, the free action  Eq.~(\ref{S0}) it is diagonalized by the unitary matrix
\be
\hat \L (\bs k) = \begin{pmatrix}
\L(\bs k) & 0 \\
0 & \mc T^{-1}\L(\bs k)
\end{pmatrix} = \left[\L_0(\bs k) + \L_z(\bs k) \tau^z\right]\,.
\ee\
Here $\L(\bs k)$ is the $4\times 4$ matrix which diagonalizes $H_e(\bs k)$ (i.e. $\L^\dag (k) H_e(k)\L(k) = \mrm{diag}\left\{ \e_k,\e_k,-\e_k,-\e_k\right\}$), $\L_0 (\bs k)\equiv{1\over 2}\left(1 + {\mc T ^{-1}}\right)\L(\bs k)$, $\L_z (\bs k)\equiv{1\over 2}\left(1 - {\mc T^{-1}}\right)\L(\bs k)  $ and $\tau^i$ represent Pauli matrices in Nambu space.
The free part of the action, written in the band basis is therefore
{ \be
\mc S_0 = {1\over 2}\sum_{\bs k,\w} C_{\bs k\a}^\dag\left[ -i\w +\left( D_{\a \a}(\bs k) -\e_F\right) \tau^z\right] C_{\bs k\a}
\ee }
where $\a$ runs over spin and band basis,
\[C_k=\hat\L^\dag(k)\begin{pmatrix} \psi_{\bs k} \\   \psi_{-\bs k}^\dag \end{pmatrix}\,\;\;\;\mrm{and}\;\;\; \Psi_{\bs k} =  \hat\L (\bs k) \begin{pmatrix} c_{\bs k} \\    c_{-\bs k}^\dag \end{pmatrix}\,.\]
\subsubsection{Interaction part}
We now turn to write the Coulomb interaction in the Nambu basis. We start from
\be
\mc S _I = {T\over 2L^3} \sum_{\bs q,\bs k,\bs k'}\sum_{\nu,\w,\w'} V(i\nu,\bs q) \left({{1\over 2} \, \Psi_{\bs k+\bs q}^\dag  \tau^z\Psi _k }  \right)\left( {1\over 2} \Psi_{\bs k'-\bs q}^\dag  \tau^z\Psi _{\bs k'}\right)
\ee
where $L^3$ is total volume.

It is useful to note that the density in Nambu space transforms to the band basis as follows
\[{\Psi_{\bs k+\bs q}^\dag  \tau^z\Psi _{\bs k} } = { \left[C_{\bs k+\bs q}^\dag\hat \L^\dag(\bs k+\bs q)\right]  \tau^z\left[\hat\L(\bs k) C _{\bs k}\right] } =  C_{\bs k+\bs q}^\dag \, M(\bs k,\bs k+\bs q) \tau^z   \, C _{\bs k}  \]
where
\[M  (\bs k,\bs p)=\L_0^\dag(\bs p) \L_0(\bs k)+\L_z^\dag(\bs p) \L_z(\bs k) = \L^\dag(\bs p) \L(\bs k) \]
Therefore, the interaction assumes the following form when written in the band basis
{
\be
\mc S _I = {T\over 8L^3} \sum_{\bs q, \bs k, \bs k'}\sum_{\nu,\w,\w'}\mc Q_{\a\b;\g\d} (i\nu, \bs k ,\bs k', q)  \;C^\dag _{\bs k+\bs q\a}   \tau^z   C_{\bs k\b}   \;C^\dag _{\bs k'-\bs q\g} \tau^{z} C_{\bs k'\d} \label{SI}
\ee}
where
\be
\mc Q _{\a\b;\g\d} (i\nu, \bs k ,\bs k', \bs q) = V(i\nu,q)M_{\a \b} (\bs k,\bs k+\bs q) M _{ \g\d}(\bs k',\bs k'-\bs q)  \label{Gamma_def}\ee
is a rank 4 tensor which obeys the equality
\be
\mc Q _{\a\b;\g\d} (i\nu, \bs k ,\bs k', \bs q) = \mc Q _{\g\d;\a\b} (i\nu, \bs k' ,\bs k, -\bs q) \label{Gamma}
\ee
We also note that $\G$ is an even function of $i \nu$.

\subsubsection{Projection to the occupied bands}
Since only two bands are occupied in each pocket we restrict the analysis to those bands. In the band basis this a trivial task: We simply restrict the sum over band indices $\a,\b,\g,\d$ to the hole bands. Therefore they now become indices running over {\it two} hole bands which are related to each other by TRS and are denote by $\a = \pm$.

\subsection{Derivation of the gap equation}
We now turn to derive the Eliashberg theory for superconductivity due to the interaction  Eq.~(\ref{Coulomb}). We first introduce the definition of the self-energy
\be
\Sigma(i\w,k) = \mc G_0^{-1}(i\w,\bs k) - \mc G^{-1}(i\w,\bs k)
\ee
where $\mc G_0^{-1}(i\w,\bs k) =  -i\w +\e(\bs k) \tau^z$ is the bare Green's function in the band basis and $\mc G(i\w,\bs k)$ is the dressed one. The self energy is then given by
{\begin{align}
\Sigma_{ \b\g}(i\w,\bs k) &= -{T \over 2 L^3}\sum_{\w',\bs k'}\sum_{\a,\d} \tau^z \mc G_{\d\a}(i\w',\bs k') \tau^{z} \left[ \mc Q_{\a\b;\g \d}(i\w-i\w',\bs k,\bs k',\bs k'-\bs k) +\mc Q_{\g\d;\a \b}(i\w'-i\w,\bs k',\bs k,\bs k-\bs k') \right] \\ &= -{T \over   L^3}\sum_{\w',\bs k'}\sum_{\a,\d} \tau^z \mc G_{\d\a}(i\w',\bs k') \tau^{z}\, \mc Q_{\a\b;\g \d}(i\w-i\w',\bs k,\bs k',\bs k'-\bs k)\nn
\end{align}}
where the two terms on the r.h.s. of the first line come from  two possible contractions of the interaction Eq.~(\ref{SI}) with $q = \bs k'-\bs k$ (note that here each one of these contractions can be taken in two equivalent ways due to the fact that we have artificially doubled the the number of fields when going to Nambu space). In the latter diagram one needs to interchange $\a,\b$ with $\g,\d$ and $k$ with $k'$. In the transition to the second line we have used Eq.~(\ref{Gamma}).
\end{widetext}

For simplicity we will neglect dispersion and mass corrections (These are typically important for extremely accurate calculation of the gap in the limit of intermediate coupling strength). In this case we have
\[\Sigma (i\w,\bs k) =\begin{pmatrix}
0 & \hat \D(i\w,\bs k)\\
\hat \D  (i\w,\bs k)  &0
\end{pmatrix}\]
where $\hat \D(i\w,\bs k) =\D_0  (i\w,\bs k) + \bs d (i\w,\bs k)\cdot \bs \s $.

For simplicity we assume the gap function is well defined under inversion, that is $\hat \D (i\w,-\bs k) = \pm \hat \D (i\w,\bs k)$ (Note that in Bismuth does ot have inversion symmetry and therefore, in general, one needs to consider a more generic gap function). We can consider two distinct cases: even parity, where $\D_0 (i\w,\bs k)\ne 0 $ and $\bs d (i\w,\bs k) = 0$ or odd parity, where $\D_0 (i\w,\bs k)= 0 $ and $\bs d (i\w,k) \ne 0$. In this case we can write
\[\tau^z\mc G(i\w,\bs k)\tau^z =- {i\w +\e(\bs k)\tau^z-\hat \D \tau^x\over \w^2 + \e^2(\bs k) + |\hat\D(i\w,\bs k)|^2}\]
where $|\hat\D(i\w,\bs k)| = |\bs d(i\w,\bs k)|$ for odd parity and is trivially defined for even parity. Therefore
\begin{widetext}
\be
\hat \D_{\b\g}(i\w,\bs k) = -{T \over   L^3}\sum_{\w',\bs k'} {M_{ \g\d}(\bs k',\bs k)\hat \D_{ \d \a}(i\w',\bs k') M_{ \a\b}(\bs k,\bs k')\over {\w'}^2 + \e^2(\bs k') + |\hat \D(i\w',\bs k')|^2} V(i\w -i\w',\bs k'-\bs k) \label{eliashberg_sup}
\ee

\subsection{Estimation of the transition temperature}
To estimate the transition temperature we linearize Eq.~(\ref{eliashberg}), i.e. we neglect the $\D$ dependence in the denominator. We also consider, for simplicity, a $s$-wave gap function.  In this case we have
\[M_{ \g\d}(\bs k',\bs k)\hat \D_{ \d \a}(i\w',\bs k') M_{ \a\b}(\bs k,\bs k')= {1\over 2} \left(1 + {\bs k \cdot \bs k' \over k^2} \right)\,\d_{\b\g}\]
Note that here we have normalized the matrices such that $\mrm{Tr}\left[ \hat \D^\dag(k)\hat \D (k) \right] = 1$.
Taking the sum over momentum to an integral and performing the integral over the solid angle we obtain
\be
 \D(i\w,x) = -{ \rho T_c \over  4\e_F   }\sum_{\w'} \int_0 ^{\infty} dx' x'^2 { \D(i\w',x') \over \left({\w' \over \e_F}\right)^2 + \left( x'-1\right)^2} V_{s}(i\w - i\w', x ,x')  \label{eliashberg_linear}
\ee
where $V_s$ is given by
\be
V_s(i\w-i\w',x,x') \equiv {1\over 2}\int_{-1} ^1 d\g \left(1+\g\right) V\left (i\w-i\w', k_F\sqrt{x^2+x'^2 - 2 x x' \g} \right)
\ee

\end{widetext}

\section{Ruling out the phonon mechanism for superconductivity}\label{app:phonon}
Throughout this paper  we have only considered Coulomb interactions, neglecting all possible local interactions. The argument was that the density of states in bismuth is too low, such that local interactions are negligible. In this section we elaborate on this point.

Let us first consider the standard electron-phonon coupling to the longitudinal acoustic phonon considered for conventional superconductors~\cite{de1989superconductivity}
\be
H_{el-ph}^{LA} = \sum_{k,q} {(-iq)D \over \sqrt{2 \rho_m \w_q} }\left(b_q + b^\dag_{-q} \right)\psi_{k+q} ^{\dag}\psi_{k}\,,
\ee
where $\w_q = v_s q$ is the dispersion of the phonon branch, $v_s = 1790 \, \mrm{m/sec}$ is the sound velocity, $\rho_m = 10 \, \mrm{g/cm^3}$ is the mass density and $b_q,\, b^\dag _q$ are the phonon creation and annihilation operators, respectively. $D$ is the deformation potential, which is typically of order $5\, \mrm{eV}$ and in Bismuth has been estimated to be $8 \,\mrm{eV}$, at most~\cite{katsuki1969calculation}.
The resulting phonon-mediated interaction in Matsubara space is then given by
\be
V_{LA}(i\w,q) = {D^2 q^2 \over \rho_m  }{1\over \w^2 + \w_q^2 }
\ee
From this, we can estimate the coupling strength by taking the zero frequency limit of the interaction times the density of states in Table \ref{tab:parameters}. We find that
\be
\lambda_h = {R D^2 \over \rho_m v_s^2} \approx0.004 \;\;;\;\; \lambda_e = {\rho D^2 \over \rho_m v_s^2} \approx0.001 \label{lambda-phonon}
\ee
which corresponds to a transition temperature, $T_c = \Theta_D \exp\left[-1/\lambda\right]$, which is much lower than $10^{-100}\,\mrm K$.
Since  Eq. \eqref{lambda-phonon} is independent on $q$ we may is also roughly estimates the coupling to the longitudinal optical branches by extrapolating $q$ to the zone boundary.

It is important to note two additional factors that reduce the effectiveness of the attractive interaction Eq. (\ref{lambda-phonon}). First, we did not take into account the matrix element effect Eq.~\eqref{form factor}, which will further reduce $\lambda$ in the electron bands. The second is that we have neglected any repulsion coming from the Coulomb interaction \eqref{Coulomb}, which will be poorly screened in the low density limit and will completely overwhelm the small attraction coming from these phonons.

We also find evidence to rule out the phonon mechanism in bismuth based on recent density functional theory (DFT) calculations~\cite{meinert2016unconventional}. In this study the author studied superconductivity in YPtBi, which is characterized by a density of $2\times 10^{18}\,\mrm{cm}^{-3}$ and a density of states $5\times 10^{19}\mrm{eV}^{-1} \mrm{cm}^{-3}$ (Comparing with Table \ref{tab:parameters} the density of states is comparable to that in bismuth). He ignored the polar nature of the phonons and used a mesh of $5\times 5\times 5$ and $10\times 10\times 10$ for the phonon and electron dispersion, which is clearly too small because the grid must be dense on the scale of $k_F$, nevertheless he found the coupling constant is $\lambda  = 0.02 \pm 0.02$. Since the electron phonon coupling in YPtBi is not anomalously small we conclude that these calculations rule out the possibility of non-polar phonon superconductivity at such low density, in agreement with the early results of Ref.~\cite{Gurevich1962}.

We can also estimate the coupling constant $\lambda$ based on the superconducting states observed in amorphus bismuth below $T_a = 6 K$~\cite{Shier1966}.
The electronic density is $\sim2\times 10^{22}\,\mrm{cm}^{-3}$, thus we estimate the density of states $\rho_a \approx 1.6\times 10^{21}\mrm{eV}^{-1}\mrm{cm}^{-3}$ and the phonon mediated interaction to be $V_0 \approx 1/\left( \rho_a \log {\Theta_D / T_a} \right)\approx 3\,\mrm{eV}\, \mc V_{uc}$, where $\mc V_{uc}$ is the unit cell volume. Assuming that the short range physics of the electron phonon coupling does not change dramatically in crystalline bismuth we estimate $\lambda_e = \rho V_0 \approx 0.002$ and $\lambda_h = \rho V_0 \approx 0.007$. Again, too small to be relevant for superconductivity.

Finally, it is also important to point out that in spite of the very low $T_c$ in bismuth the coupling constant can not be smaller that $\lambda = 0.08$. Inverting the density of states and assuming a general local interaction of the form
\be
H_{BCS} = -V_0\sum_{k,k'}\psi^\dag_k \psi^\dag_{-k}\psi_{-k'} \psi_{k'}\label{BCS-int}
\ee
and taking $V_0 = (\rho\log\Theta_D / T_c)^{-1}$ we find that the phonon mediated interaction must be at least $V_0 \approx 135\,\mrm{eV} \mc V_{uc} $. Thus, when interpreting the attractive interaction in bismuth as local, with a typical crystalline length scale, it takes an unphysically large value indicting that superconductivity does not come from local interactions.

\section{Matrix element modifications of the BCS formula for $T_c$} \label{andersons}
The additional form factor (\ref{form factor}) has an important effect on superconductivity, regardless of the pairing mechanism. To see this, let us consider the constant attractive interaction that BCS considered in their original paper (i.e. Eq.\eqref{BCS-int}, which corresponds to $V(i\w-i\w',\bs k -\bs k') = -V_0 \Theta (\w_D-|\w|)$). Without spin-orbit coupling the $M(\bs k,\bs k')$ matrices become trivial and Eq.~(\ref{eliashberg}) reduces to (for $s$-wave pairing)
 \be
 {1\over \rho V_0} =   \int _{T_{c}} ^{\w_D} {d\w  \over |\w|}\label{eliashberg3}
 \ee
where $\rho = k_F^2 / 2\pi^2 v$ is the electronic density of states per pocket and spin. Note that we have made the assumption $\w_D \ll \e_F$.
The corresponding transition temperature is given by Eq.(\ref{BCS}).
In contrary, within the Dirac dispersion, where spin orbit coupling is present, we find two crucial differences:

 \noindent
 (i) First we find that the odd parity pairing channel [of the form $\hat \D (\bs k) = - \hat \D(-\bs k)$], may also have a finite transition temperature in spite of the structureless interaction assumed by BCS. This is contracts to the case of no spin-orbit coupling, where one finds that the transition temperature to an odd-parity state is strictly zero.
 This origin for this difference is the matrices $M_{\bs k ,\bs k'}$ in Eq. (\ref{eliashberg}) which encode the channel's parity into non-trivial momentum dependent form factors. Note that for odd-parity pairing $\D(i\w,\bs k)$ is proportional to a sum of the Pauli matrices and therefore the factor (\ref{form factor}) takes a different form.

\noindent
 (ii) Second, we find that when averaging (\ref{form factor}) over the solid angle between $\bs k $ and $\bs k'$ the second term in the l.h.s. averages to zero. When $\D_{bg} = 0$ Eq.~(\ref{form factor}) gives a factor of $1/2$ reduction of the coupling constant. For finite $\D_{bg}$ we find (for $s$-wave pairing)
\be
 T_{c} = \w_D \exp\left\{ {-{1\over \rho V_0} \left[{{2}+ 2\left({v k_F \over \D_{bg}}\right)^2  \over 2+\left({v k_F \over \D_{bg}}\right)^2 }\right] }  \right\} \label{violation}
 \ee
Thus, in the parabolic limit $v k_F \ll \D_{bg}$ the transition temperature reduces to Eq.~(\ref{BCS}). However, in the relativistic limit $v k_F \gg \D_{bg}$ the coupling constant is reduced by a factor of $1/2$ and consequently $T_c$ is reduced exponentially. Note that here we have assumed a constant density of states. Thus the parameter $\D_{bg} / v k_F$ continuously tunes between Eq. (\ref{BCS}) and the much suppressed transition temperature $T_c = \w_D \exp\left[-2/\rho V_0 \right]$.

The latter is in sharp contradiction to the conventional wisdom based on Anderson's~\cite{anderson1959theory} notion of pairing time reversed states which states that any time-reversal-symmetric perturbation that does not modify the density of states also does not modify the transition temperature. The origin of the contradiction is the projection to the occupied bands, which was not considered as a possibility in Ref.~\cite{anderson1959theory}. Anderson used the completeness of eigen states to prove that the transformation from one eigen basis to the other does not modify the matrix element in the $s$-wave channel. However, the action of projection violates the completeness of the basis and allows to get an overlap smaller than one.

A similar matrix element effect has been discussed in detail in Ref.~\cite{Savary}, where superconductivity in a quadratic band touching point has been considered. It is found that the suppression of the $s$-wave channel is maximal, i.e. the form factor (\ref{form factor}) is equal to $1/2$ as long as the band touching point is not gapped. Note that the same applies to Dirac semi-metals, namely in the case of $\D_{bg} = 0$ the form factor (\ref{form factor}) becomes equal to $1/2$ after angular averaging.

\end{appendices}

\end{document}